\documentclass[12pt]{article}
\usepackage{epsfig,cite}
\usepackage{amsmath}
\usepackage{amssymb}
\usepackage{color}
\usepackage{graphicx}
\usepackage{verbatim,mathrsfs,subfigure,feynmf}
\usepackage{url}
\usepackage{ytableau}
\usepackage{multirow}
\include{epsf}

\newcount\timecount
\newcount\hours \newcount\minutes  \newcount\temp \newcount\pmhours
\hours = \time
\divide\hours by 60
\temp = \hours
\multiply\temp by 60
\minutes = \time
\advance\minutes by -\temp
\def\hour{\the\hours}
\def\minute{\ifnum\minutes<10 0\the\minutes
            \else\the\minutes\fi}
\def\clock{
\ifnum\hours=0 12:\minute\ AM
\else\ifnum\hours<12 \hour:\minute\ AM
      \else\ifnum\hours=12 12:\minute\ PM
            \else\ifnum\hours>12
                 \pmhours=\hours
                 \advance\pmhours by -12
                 \the\pmhours:\minute\ PM
                 \fi
            \fi
      \fi
\fi
}

\def\monthname{\relax\ifcase\month 0/\or January\or February\or
   March\or April\or May\or June\or July\or August\or September\or
   October\or November\or December\else\number\month/\fi}

\def\bold#1{\setbox0=\hbox{$#1$}%
     \kern-.025em\copy0\kern-\wd0
     \kern.05em\copy0\kern-\wd0
     \kern-.025em\raise.0433em\box0 }

\newcommand\SH[1]{{\color{black} {#1}}}

\def\beq{\begin{equation}}
\def\eeq{\end{equation}}

\def\ga{\mathrel{\raise.3ex\hbox{$>$\kern-.75em\lower1ex\hbox{$\sim$}}}}
\def\la{\mathrel{\raise.3ex\hbox{$<$\kern-.75em\lower1ex\hbox{$\sim$}}}}
\def\gev{{\rm \, Ge\kern-0.125em V}}
\def\tev{{\rm \, Te\kern-0.125em V}}
\def\gyr{{\rm \, G\kern-0.125em yr}}

\def\gappeq{\mathrel{\rlap {\raise.5ex\hbox{$>$}}
{\lower.5ex\hbox{$\sim$}}}}
\def\lappeq{\mathrel{\rlap{\raise.5ex\hbox{$<$}}
{\lower.5ex\hbox{$\sim$}}}}
\def\Toprel#1\over#2{\mathrel{\mathop{#2}\limits^{#1}}}

\def\m12{m_{1\!/2}}

\def\bea{\begin{eqnarray}}
\def\eea{\end{eqnarray}}

\usepackage{lscape}

\def\beq{\begin{equation}}
\def\eeq{\end{equation}}

\newcommand{\1}{\mbox{1}\hspace{-0.25em}\mbox{l}}

\begin{document}\begin{titlepage}
\pagestyle{empty}
\begin{flushright}
{\tt KCL-PH-TH/2026-08}, {\tt CERN-TH-2026-036}  \\
{\tt UMN-TH-4522/26, FTPI-MINN-26/06} \\
\end{flushright}

\vspace{1cm}
\begin{center}{\bf \large{Planck-Scale Effects on Nucleon Decay \\
\vspace{2mm}
in Minimal Supersymmetric SU(5)}}\\
\vskip 0.4in
{\bf John~Ellis}$^{a}$,
{\bf Jason~L.~Evans}$^{b,c}$,
{\bf Shihwen Hor}$^{b}$,
{\bf Natsumi Nagata}$^{d}$ {\small \&}
{\bf Keith~A.~Olive}$^{e}$
\vskip 0.2in
{\small
{\em $^a$Theoretical Particle Physics and Cosmology Group, Department of
  Physics, King's~College~London, London WC2R 2LS, United Kingdom;\\
Theoretical Physics Department, CERN, CH-1211 Geneva 23,
  Switzerland;\\
National Institute of Chemical Physics and Biophysics, R\"{a}vala 10, 10143 Tallinn, Estonia}\\[0.2cm]
  {\em $^b$Tsung-Dao Lee Institute, Shanghai Jiao Tong University, Shanghai 200240, China}\\[0.2cm] 
  {\em $^c$American Physical Society, Hauppauge, New York, USA}\\[0.2cm] 
  {\em $^d$Department of Physics, University of Tokyo, Bunkyo-ku, Tokyo
 113--0033, Japan}\\[0.2cm] 
{\em $^e$William I. Fine Theoretical Physics Institute, School of
 Physics and Astronomy,\\ University of Minnesota, Minneapolis, MN 55455,
 USA}}
 
\vspace{1cm}
{\bf Abstract}
\end{center}
\baselineskip=18pt \noindent
{\small
We examine the impact on the phenomenology of the minimal supersymmetric SU(5) Grand Unified Theory (GUT) of dimension-5 operators with coefficients suppressed by the Planck mass scale, with particular emphasis on predictions for nucleon decay. We incorporate dimension-5 operators in both the Higgs sector and the Yukawa interactions in the theory, and take account of the constraints from gauge coupling measurements, the mass of the Higgs boson, fermion masses and the cold dark matter density. We consider two scenarios for soft supersymmetry breaking: the constrained minimal supersymmetric extension of the Standard Model (CMSSM) and the Non-Universal Higgs Model (NUHM). We present predictions for the nucleon decay modes $p \to \pi^0 e^+, \pi^0 \mu^+, K^+ \bar \nu, \pi^+ \bar \nu$, $K^0 e^+, K^0 \mu^+$ and $n\to \pi^0 \bar \nu$, $\pi^- e^+, K^0 \bar \nu$, which we compare with both the present experimental sensitivities and those projected for the JUNO and Hyper-Kamiokande experiments. We find that these experiments may have interesting possibilities for discovering several of these decay modes.}


\vfill

\leftline{March 2026}
\end{titlepage}

\section{Introduction}
\label{sec:intro}
The advent of a new generation of large underground, low-background neutrino detectors, represented by JUNO \cite{JUNO:2022qgr}, Hyper-Kamiokande \cite{Hyper-Kamiokande:2018ofw}, and DUNE~\cite{DUNE:2020fgq} motivates theoretical re-examinations of nucleon decays in Grand Unified Theories (GUTs) \footnote{For a recent overview of future searches for baryon number violation in neutrino experiments, see \cite{Dev:2022jbf}.}. In this paper we look again at nucleon decays in the minimal supersymmetric SU(5) GUT~\cite{Dimopoulos:1981zb, Sakai:1981gr}, taking into account the possible effects of dimension-5 operators~\cite{Weinberg:1981wj, Sakai:1981pk} generated by Planck-scale dynamics that might appear in both the Higgs sector of the model and the fermion Yukawa interactions. The general expectation in the absence of such interactions is that $p \to K^+ \bar \nu$ and $n \to K^0 \bar \nu$ would be the dominant nucleon decay modes~\cite{Dimopoulos:1981dw, Ellis:1981tv}, but this may no longer be the case when dimension-5 interactions are taken into account, as we discuss in this paper.

It has long been realized that, because the GUT scale inferred from proton stability and renormalization group (RG) analyses cannot be much smaller than the Planck scale \cite{Goto:1998qg,Murayama:2001ur}, the dynamics at this scale may impact GUT phenomenology. Two important examples are the possible contributions of dimension-5 Planck-scale interactions to the masses of matter fermions, and modifications of the GUT gauge-coupling unification conditions by dimension-5 operators in the GUT Higgs and gauge sectors \cite{Ellis:1979fg,Panagiotakopoulos:1984wf,Bajc:2002pg}. Their effects are particularly important in supersymmetric GUTs, since the unification scale in such models is estimated to be significantly higher than in simple non-supersymmetric GUT models.

The possible effects of dimension-5 operators on the GUT gauge-coupling unification conditions are two-fold  \cite{Ellis:2016tjc,Ellis:2017djk,Ellis:2019fwf,Ellis:2020mno}. On the one hand, they may modify the relations between the masses of the different components of the GUT Higgs bosons and gauge bosons, modifying the RG running of the gauge couplings in the approach to the unification scale. In addition, they may modify the gauge coupling unification conditions themselves.

There are also two possible types of effects of dimension-5 operators on nucleon decay modes and branching ratios \cite{Emmanuel-Costa:2003szk,Ellis:2019fwf,Ellis:2020mno,Babu:2020ncc,Hamaguchi:2024ewe}. One originates in the GUT gauge-Higgs sector, due to changes in the relative importance of gauge- and Higgs-mediated decay amplitudes. The other originates in the fermionic sector, as dimension-5 operators that contribute to light fermion masses may modify the favoured nucleon decay branching ratios.

The layout of our paper is as follows. In Section~\ref{sec:model} we first set out our formulation of the dimension-5 operators in the Higgs-gauge sector of the minimal supersymmetric SU(5) GUT model that we consider. We then calculate the corrections that they introduce in the GUT Higgs vacuum expectation value (vev) that breaks SU(5)$\to$SU(3)$\times$SU(2)$\times$U(1), keeping terms up to second order in supersymmetry breaking masses. We then calculate the mass spectrum of the GUT-scale components of the $\mathbf{\bar 5}$ and $\mathbf{24}$ Higgs multiplets, keeping track of their relation to the GUT gauge bosons. We then display the RG equations for the U(1), SU(2) and SU(3) gauge bosons and the perturbed gauge coupling unification conditions. We also display the corrections to the gaugino masses. Finally we consider the dimension-5 operator contributions to the
Yukawa couplings of the colored Higgs and uncolored Higgs bosons. As we discuss, whilst the couplings to the uncolored Higgs are constrained by the matter fermion masses, there is ambiguity in the couplings to the colored Higgs, which allows the possibility of suppressing proton decay into $K^+ \bar \nu$ and allowing $p \to \pi^0 e^+$ to become the dominant decay mode. A similar mechanism was discussed in Ref.~\cite{Dorsner:2024seb}. 

We turn in Section~\ref{sec:pdecay} to the details of how the lifetimes and decay modes of nucleons are affected, considering the proton decay modes
$p \to \pi^0 e^+, \pi^0 \mu^+, K^+ \bar \nu, \pi^+ \bar \nu$, $K^0 e^+, K^0 \mu^+$ and the neutron decay modes $n\to \pi^0 \bar \nu$, $\pi^- e^+, K^0 \bar \nu$, which are related by isospin to the corresponding proton decay modes.
Numerical results are presented graphically in Section~\ref{sec:results}. We show the sensitivities of the proton lifetime in different modes to several GUT couplings and compare these sensitivities to those stemming from Planck-suppressed operators. We consider two representative benchmark points in the constrained minimal supersymmetric extension of the Standard Model (CMSSM). We also consider a simple extension, namely the non-universal Higgs model (NUHM).
Finally, we present and discuss our conclusions in Section~\ref{sec:conclusion}.

\section{The Model}
\label{sec:model}

In this work, we examine a generalized version of the minimal supersymmetric SU(5), analyzing in detail the effect of Planck-suppressed dimension-5 operators on the predictions of minimal supersymmetric SU(5) for baryon decays. 
The superpotential can be split as $W=W^H_{\rm eff}+W^{\Delta h}_{\rm eff}$, where $W^{\Delta h}_{\rm eff}$ is responsible for the Yukawa interactions and $W^H_{\rm eff}$ contains Higgs interactions.
The superpotential of the Higgs sector we consider (following the notation of \cite{Ellis:2010jb,Ellis:2016qra,Ellis:2016tjc,Ellis:2019fwf,Ellis:2020mno})  is 
\begin{eqnarray}
W^H_{\rm eff} &=& \mu_\Sigma {\rm Tr}\left(\Sigma^2\right) +\frac{\lambda'}{6} {\rm Tr}\left(\Sigma^3\right)+\frac{\kappa_1}{4M_P}{\rm Tr}\left(\Sigma^4\right)+\frac{\kappa_2}{4M_P}\left({\rm Tr}\left(\Sigma^2\right)\right)^2 +\mu_H H\bar H +\lambda H\Sigma \bar H \nonumber\\ 
&+&\frac{c_1}{M_P}H\Sigma^2\bar{H} +\frac{c_2}{M_P}H\bar{H} {\rm Tr} \left( \Sigma^2 \right) + \frac{c_3}{M_P}(H\bar{H})^2+\frac{c}{M_P} {\rm Tr}\left(\Sigma {\mathcal W} {\mathcal W}\right) ~.
\end{eqnarray}
The SU(5) GUT gauge group is spontaneously broken to the Standard Model (SM) gauge group by the vev of the Higgs adjoint, a ${\bf 24}$ chiral superfield,
$\Sigma \equiv \sqrt{2}\Sigma^A T^A$, 
where the generators of SU(5) are normalized to 1/2, i.e., ${\rm Tr}(T^aT^b)=\delta^{ab}/2$.  
Higgs fields in the ${\bf 5}$ and $\overline{\bf 5}$
representations are denoted by  $H$ and $\overline{H}$. In the last term, ${\cal W}\equiv  {\cal W}^A T^A$ denotes the
superfields corresponding to the field strengths of the SU(5) gauge vector bosons
${\cal V} \equiv {\cal V}^A T^A$. 

In addition, as with the dimension-5 superpotential operators, K\"{a}hler-type dimension-5 operators are generally also allowed. For example, the following operators generated from K\"{a}hler terms could also be considered,\footnote{For completeness, we note that 
there are also K\"ahler corrections of the form 
\begin{equation}
     \Delta K_{\rm eff}= \frac{d^{ij}_{\Delta h, 1}}{M_P} \epsilon_{\alpha\beta\gamma\delta\zeta}
 \Psi_i^{\alpha\beta} \Psi^{\gamma\delta}_j (\overline{H}^*)^\zeta + \frac{d^{ij}_{\Delta h, 2}}{M_P} \Psi_i^{\alpha\beta} \Phi_{j \alpha}
({H}^*)_\beta + \mathrm{h.c.} \, .\nonumber
\end{equation}
Once the $F$-terms of $H$ and $\bar{H}$ are integrated out, $F_H = - \mu_H^* \overline{H}^* + \dots $ and $F_{\bar{H}} = - \mu_H^* H^* + \dots $, these terms give $\mathcal{O} (\mu_H / M_P)$ corrections to the Yukawa couplings. Since these corrections can be absorbed by the redefinition of $h_{\mathbf{10}}$ and $h_{\overline{\mathbf{5}}}$, they do not affect our discussion.  
}
 
\begin{align}
    \Delta K_{\rm eff}= \frac{c_{\Sigma\Sigma}}{M_P}{\rm Tr} (\Sigma^* \Sigma^2) + \frac{c_{\bar{H}\Sigma H}}{M_P}\bar{H}  \Sigma^* H  +\sum_{F=\Psi,\Phi,H,\bar{H}} \frac{c_{F\Sigma}}{M_P}F^*\Sigma F \, +{\rm h.c. }~.
\end{align}
The first term will contribute to the $F$-term of $\Sigma$ while the second term may contribute to MSSM value of $\mu$. The last term in the above equation modifies the wave functions of the field $F$ below the GUT scale, but the effect is insignificant for our analysis and is neglected in what follows.

In addition to the superpotential of the Higgs sector, there are also higher-dimensional terms in the superpotential of the Yukawa sector. 
This part of the superpotential can be written as
\begin{align}
  W_{\rm eff}^{\Delta h} &= 
\left(h_{\bf 10}\right)_{ij} \epsilon_{\alpha\beta\gamma\delta\zeta}
 \Psi_i^{\alpha\beta} \Psi^{\gamma\delta}_j H^\zeta +
 \left(h_{\overline{\bf 5}}\right)_{ij} \Psi_i^{\alpha\beta} \Phi_{j \alpha}
\overline{H}_\beta \nonumber \\  
  & + \frac{c_{\Delta h, 1}^{ij}}{M_P} \Psi_i^{\alpha\beta} \Phi_{j\alpha}
\Sigma^\gamma_{~\beta} \overline{H}_\gamma  +\frac{c_{\Delta h,2 }^{ij}}{M_P} \Phi_{i \alpha}
  \Sigma^\alpha_{~\beta} \Psi^{\beta\gamma}_j \overline{H}_\gamma\nonumber \\
&
+\frac{c_{\Delta h, 3}^{ij}}{4M_P}
 \epsilon_{\alpha\beta\gamma\delta\zeta}
 \Psi_i^{\alpha\beta} \Psi^{\gamma\xi}_j \Sigma^\delta_{~\xi} H^\zeta 
+ \frac{c_{\Delta h, 4}^{ij}}{4M_P}
\epsilon_{\alpha\beta\gamma\delta\zeta}
 \Psi_i^{\alpha\beta} \Psi^{\gamma\delta}_j \Sigma^\zeta_{~\xi} H^\xi   \, .
\label{eq:weffdelh}
\end{align}
The multiplets ${\Phi}_i$ in Eq.~\eqref{eq:weffdelh} are  $\bf{\overline{5}}$ representations containing the left-handed SM matter fields
$\overline{D}_i$ and ${L}_i$, and the ${\Psi}_i$ are $\bf{10}$ representations of SU(5) containing the left-handed
${Q}_i$, $\overline{U}_i$, and $\overline{E}_i$,
where the index $i = 1,2,3$ denotes the
generations. The first line of Eq.~\eqref{eq:weffdelh} provides the standard Yukawa couplings of the matter fields to the Higgs 5-plets. 
The last two lines correspond to Planck-suppressed operator contributions to the Yukawa couplings of the Higgs and colored Higgs. 
These terms affect the relation between the GUT Yukawa couplings, $h_{\bf 10}$ and $h_{\overline{\bf 5}}$ and the Standard Model couplings, as we describe in more detail below. 
An analysis including $\kappa_1$, $\kappa_2$, and $c_{\Delta h,2}$, $c_{\Delta h,1}$, $c_{\Delta h,3}$, $c_{\Delta h,4}$, was performed in \cite{Emmanuel-Costa:2003szk}, and~\cite{Babu:2020ncc} also considered the first three of these couplings.

Finally,  it is also possible in this context to include additional Planck-scale dimension-5 B- and L-violating operators:
\begin{align}
    W_{\rm eff}^{L\!\!\!/} &=\frac{c_{L\!\!\!/}^{ij}}{M_P}\Phi_{i\alpha}\Phi_{j\beta}H^\alpha H^\beta~, 
    \label{eq:Lviolating} \\ 
    W_{\rm eff}^{B\!\!\!\!/L\!\!\!/} &=\frac{c_{B\!\!\!\!/L\!\!\!/}^{ijkl}}{M_P}\epsilon_{\alpha\beta\gamma\delta\zeta}\Psi_i^{\alpha\beta}\Psi_j^{\gamma\delta}\Psi_k^{\zeta\xi}\Phi_{l\xi}~.
    \label{eq:Bviolating}
\end{align}
Eq.~\eqref{eq:Lviolating} can provide Majorana masses for left-handed neutrinos. However, these masses are smaller than is required by neutrino oscillation observations. Neutrino masses of the appropriate order of magnitude can be generated with the inclusion of right-handed neutrinos.
Eq.~\eqref{eq:Bviolating} includes $QQQL$ and $\bar{U}\bar{D}\bar{U}\bar{E}$ operators that can induce rapid proton decay. However, these operators are flavor-dependent, and the flavor symmetry that suppresses mixing among the Standard Model quarks could constrain these operators as well, suppressing operators involving the first two generations. As can be seen below, the most naive implementation of this method of suppression is inconsistent with the magnitudes of the Planck-suppressed operators we postulate to suppress proton decay.  
We may instead invoke some other symmetry to suppress $W_{\rm eff}^{B\!\!\!\!/L\!\!\!/}$, as discussed, e.g., in Ref.~\cite{Ibanez:1991pr}. 
Finding a viable model to suppress these operators and retain the operators we consider below would require a dedicated investigation that lies beyond the scope of this work (see, e.g., Refs.~\cite{Dine:2013nga, Ibe:2024cvi, Chitose:2025bvl} for relevant discussions.).

The soft supersymmetry-breaking terms are 
\begin{eqnarray}
{\mathcal L}_{\rm soft} &&= - \left( m_{\mathbf{10}}^2 \right)_{ij} \widetilde{\psi}_i^*\widetilde{\psi}_j -\left( m_{\overline{\mathbf{5}}}^2 \right)_{ij} \widetilde{\phi}_i^*\widetilde{\phi}_j -m_H^2|H|^2-m_{\bar{H}}^2|\bar{H}|^2 -m_\Sigma^2{\rm Tr}(\Sigma^\dagger\Sigma)\nonumber\\
&& \quad\quad -\biggl[  \frac{1}{2}M_5\widetilde{\lambda}^A\widetilde{\lambda}^A + A_{\mathbf{10}}(h_{\mathbf{10}})_{ij} \epsilon_{\alpha\beta\gamma\delta\zeta} \widetilde{\psi}_i^{\alpha\beta}\widetilde{\psi}_j^{\gamma\delta}H^\zeta +A_{\overline{\mathbf{5}}}(h_{\overline{\mathbf{5}}})_{ij}\widetilde{\psi}_{i}^{\alpha\beta}\widetilde{\phi}_{j\alpha}\bar{H}_\beta \nonumber\\
&& \quad\quad\quad\quad +\mu_\Sigma  B_\Sigma {\rm Tr}\left(\Sigma^2\right) +\frac{\lambda'}{6}A_{\lambda'} {\rm Tr}\left(\Sigma^3\right)+\frac{\kappa_1}{4M_P}{\rm Tr}A_1\left(\Sigma^4\right) \nonumber\\
 &&  \quad \quad \quad \quad +\frac{\kappa_2}{4M_P}A_2\left({\rm Tr}\left(\Sigma^2\right)\right)^2+ \mu_H B_H \bar H H +\lambda A_\lambda H\Sigma \bar H +{\rm h.c.} \biggr]~, 
\label{eq:soft}
\end{eqnarray}
where $\widetilde{\psi}_i$ and $\widetilde{\phi}_i$ are the scalar components of the supermultiplets $\Psi_i~(\mathbf{10})$ and $\Phi_i~(\overline{\mathbf{5}})$ in Eq.~\eqref{eq:weffdelh}, respectively, and  $\widetilde{\lambda}^A$ are the SU(5) gauginos. Note that terms such as $\frac{c_1}{M_P}A_{c_1}H\Sigma^2 \bar{H}$, $\frac{c_2}{M_P}A_{c_2}H\bar{H} {\rm Tr} \left( \Sigma^2 \right)$ and $\frac{c_3}{M_P}A_{c_3}(H\bar{H})^2$ can also be included. They affect the MSSM Higgs $B$ term and provide additional interaction terms. However, their effects are subleading, since they can either be absorbed into other supersymmetry parameters or have a negligible effect on the analysis below, as would the $A$-terms associated with the couplings $c_{\Delta h,i}$. Therefore, we omit them from the following discussion. 

The soft mass terms are determined by imposing the following universality
conditions at the GUT scale, which we take as the initial renormalization scale before running to the electroweak scale:
\begin{align}
 \left(m_{\bf 10}^2\right)_{ij} =
\left(m_{\overline{\bf 5}}^2\right)_{ij}
&\equiv m_0^2 \, \delta_{ij} ~,
\nonumber \\[3pt]
m_H = m_{\overline{H}} = m_\Sigma &\equiv m_0 ~,
\nonumber \\[3pt]
A_{\bf 10} = A_{\overline{\bf 5}} = A_\lambda = A_{\lambda^\prime}
&\equiv A_0 ~,
\nonumber \\[3pt]
 M_5 &\equiv m_{1/2} ~.
\label{eq:inputcond}
\end{align}
Note that in Section \ref{sec:resultsNUHM}
we drop the condition $m_H = m_{\overline{H}}=m_0$, allowing both Higgs masses to be free. 
The bilinear soft supersymmetry-breaking therms $B_\Sigma$ and $B_H$ are determined from the other
parameters, as we shall see in what follows. 
These are equivalent to the boundary conditions in the Constrained Minimal Supersymmetric Model (CMSSM) (see, e.g., \cite{Ellis:2016tjc}).

Next, we examine the breaking of the GUT gauge group SU(5). First, we look for a minimum which gives the breaking pattern SU(5)$\to$SU(3)$\times$SU(2)$\times$U(1) and take the vev to be normalized as follows 
\begin{eqnarray}
\langle \Sigma \rangle =\left(V+F_\Sigma \theta^2\right){\rm diag}(2,2,2,-3,-3)~.
\end{eqnarray}
We have included an $F$-term, since the supersymmetry breaking soft masses will generate a non-zero $F$-term for $\Sigma$.
The vev can be solved for perturbatively around the supersymmetric minimum. We will need to keep terms up to second order in supersymmetry-breaking masses, so we parameterize the vev as 
\begin{eqnarray}
V=v_0+v_1 + v_2+...~,
\end{eqnarray}
where the subscript indicates how that contribution scales with the supersymmetry-breaking masses.  The solution is then~\footnote{Note that in the limit $\kappa\to 0$ one finds the conventional solution $v_0\sim 4\mu_\Sigma/\lambda'$ that we focus on in the following discussion. We note, however, that there is a second solution with $v_0\sim \infty$. 
}
\begin{eqnarray}
v_0&&=\frac{1}{4} \frac{\lambda'M_P\pm \sqrt{\lambda'^2M_P^2-32\kappa M_P\mu_\Sigma}}{\kappa} \nonumber\\
&&  \simeq 
\frac{4\mu_\Sigma}{\lambda'}+\frac{32\kappa\mu_\Sigma^2}{{\lambda'}^3M_P}~, \quad \frac{1}{2}\frac{\lambda'M_P}{\kappa}-\frac{4\mu_\Sigma}{\lambda'}-\frac{32\kappa\mu_\Sigma^2}{{\lambda'}^3M_P}  ~,\\
v_1&&=2\frac{\lambda'(A_{\lambda'}-B_\Sigma)-2\left(7\kappa_1 (A_1-B_\Sigma)+30\kappa_2 (A_2-B_\Sigma)\right) R}{\left(\lambda'-4\kappa R\right)^2}~,\\
v_2&&=
-4\frac{m_\Sigma^2}{v_0\left(\lambda'-4\kappa R\right)^2}
+2\frac{\lambda'A_{\lambda'}-4\left(7\kappa_1A_1+30\kappa_2A_2\right) R}{(\lambda'-4\kappa R)^2}\frac{v_1}{v_0}
-2\frac{\lambda'-7\kappa R}{\lambda'-4\kappa R} \frac{v_1^2}{v_0}~,
\end{eqnarray}
where 
\begin{equation}
\kappa \; = \; 7\kappa_1+30\kappa_2~, \qquad
R \; =\; \frac{v_0}{M_P}~,
\end{equation}
with $M_P = 2.4 \times 10^{18} ~{\rm GeV}$.

Because the vev of $\Sigma$ is deflected by the soft masses, the $F$-term of $\Sigma$ becomes non-zero and is given by
\begin{eqnarray}
F_{\Sigma}=\frac{1}{2} \left(\lambda'+ 4(c_{\Sigma\Sigma} - \kappa) R\right) v_0 v_1 +\frac{1}{2} \left(\lambda'-4 \kappa R\right)v_0 v_2+\frac{1}{2} \left(\lambda'-6 \kappa R\right) v_1^2 ~.
\label{eq:FSigma}
\end{eqnarray}
We can now use these expressions to compute the MSSM Higgs $\mu$ and $B$ terms~\footnote{These expressions, as well as Eq.~(\ref{eq:FSigma}), generalize those derived in \cite{Borzumati:2009hu,Ellis:2016tjc} by including the contributions of the higher-dimensional operators.},
\begin{eqnarray}
\mu &=& \mu_H -3\lambda(v_0+v_1+v_2)+ (9c_1+30c_2)Rv_0 \biggl(1+\frac{v_1+v_2}{v_0}\biggr)^2 -\frac{3c_{\bar{H}\Sigma H}}{M_P}F_{\Sigma} \nonumber\\
&=&{\mathcal O}\left(M_{SUSY}\right) ~,\label{eq:HiggMu}\\
B &=& B_H +\frac{3\lambda v_0\Delta_T}{\mu}+\frac{3\lambda}{\mu}\left(\left(B_H-A_\lambda\right)v_1+\frac{1}{2} \left(\lambda'-4 \kappa R\right)v_0 v_2+\frac{1}{2} \left(\lambda'-6 \kappa R\right) v_1^2\right)~,\label{eq:HiggBterm}
\end{eqnarray}
where
\begin{eqnarray}
\Delta_T=\frac{1}{2}\left(\lambda'-4\kappa R\right)v_1+B_H-A_\lambda={\mathcal O}\left(\frac{M_{SUSY}^2}{M_{GUT}}\right)~. \label{eq:DelT}
\end{eqnarray}
It can be shown that this parameter is scale-independent. Hence, if it is set to zero at any scale, it will stay zero at all scales~\cite{Kawamura:1994ys}. The $\Delta_T$ parameter must be tuned so that it is of order $M_{SUSY}^2/M_{GUT}$ or electroweak symmetry breaking will fail. Using this relationship allows us to get a simplified expression for $v_1$,
\begin{eqnarray}
v_1=2\frac{A_\lambda-B_H+\Delta_T}{\lambda'-4\kappa R}\simeq 2\frac{A_\lambda-B_H}{\lambda'-4\kappa R}~.
\end{eqnarray}
The Higgs $B$ term found in Eq.~\eqref{eq:HiggBterm} has the same freedom as used in Ref.~\cite{Ellis:2017djk,Ellis:2019fwf} to liberate the $B$ term from the conditions of the functional form found in Eq.~\eqref{eq:HiggBterm}. We can introduce Giudice-Masiero (GM) terms~\cite{Giudice:1988yz}
\beq\label{GM1}
\Delta K = \left(c_H H \bar{H} + c_\Sigma  \Sigma^2 + {\rm h.c.}\right) \, ,
\eeq
which give an additional contribution to $\Delta_T$ of order $M_{SUSY}^2/M_{GUT}$ and in turn a contribution of order $m_{SUSY}^2/\mu$ to the Higgs $B$ term, making electroweak symmetry breaking easier to accomplish. 
Note that the GM terms induce a shift in $\Delta_T$:
\begin{equation}
    \delta \Delta_T =\biggl( \frac{c_H}{\mu_H}-\frac{\lambda'-2\kappa R}{\lambda'-4\kappa R}\frac{c_\Sigma}{\mu_\Sigma} \biggr) 2m_{3/2}^2~.
\end{equation}
The overall shift in $B$ is ${\cal O}(M_{\rm SUSY})$ and is estimated to be 
\begin{equation}
    \frac{3\lambda V\Delta_T}{\mu} \to \biggl[ c_H\biggl( 1+\frac{(9c_1+30c_2)V^2}{\mu_HM_P}\biggr)-c_\Sigma\biggl( 12\frac{\lambda}{\lambda'}+96\frac{\lambda}{\lambda'^3}\frac{\kappa\mu_\Sigma}{M_P}\biggr)\frac{\lambda'-2\kappa R}{\lambda'-4\kappa R} \biggr] \frac{2m_{3/2}^2}{\mu}~.
\end{equation}
A similar but more detailed discussion can be found in Ref.~\cite{Ellis:2017djk,Ellis:2019fwf}.

We next calculate the GUT mass spectrum of the component fields:
\begin{align}
  H &= 
  \begin{pmatrix}
    H^1_C \\ H^2_C \\ H^3_C \\ H_u^+ \\ H_u^0 
  \end{pmatrix}
  ~, \qquad 
  \bar{H} =
  \begin{pmatrix}
    \bar{H}_{C1} \\ \bar{H}_{C2} \\ \bar{H}_{C3} \\ H_d^- \\ - H_d^0
  \end{pmatrix}
  ~, \\[3pt] 
  \Sigma & = 
  \begin{pmatrix}
    \Sigma_8&\Sigma_{(3,2)} \\
    \Sigma_{(3^*,2)} & \Sigma_3
   \end{pmatrix}
   +\frac{1}{2\sqrt{15}}
   \begin{pmatrix}
    2 \1_3 &0\\0&-3 \1_2
   \end{pmatrix}
   \Sigma_{1}~.
\end{align}
First, we look at the masses of the particles contained in $\Sigma$. With our chosen normalization, we obtain masses for the adjoint components 
\begin{eqnarray}
M_{\Sigma_3}&=& \frac{5}{2}\lambda' V-20\frac{\kappa_1V^2}{M_P}~,\nonumber\\
M_{\Sigma_8}&=& \frac{5}{2}\lambda' V+ 5\frac{\kappa_1V^2}{M_P}~,\nonumber\\
M_{\Sigma_1}&=& \frac{1}{2}\lambda' V -2\frac{\kappa V^2}{M_P}~,
\label{eq:Msig}
\end{eqnarray}
where $V$ is the vev of $\Sigma$ to all orders but is well approximated by $v_0$. The components $\Sigma_{(3,2)} $ and $\Sigma_{(3^*,2)}$ are massless Nambu-Goldstone fields and are absorbed into the massive SU(5) gauge fields.
The masses of the SU(5) gauge bosons and the color triplet Higgs fields are found to be
\begin{eqnarray}
M_X &=& 5g_5V ~,\\
M_{H_C} &=& \mu_H+2\lambda V +\frac{(4c_1+30c_2)}{M_P} V^2 \nonumber\\
 &=& 5\lambda V - \frac{5c_1}{M_P} V^2~,
 \label{eq:mhc}
\end{eqnarray}
where we have taken the Higgs doublet mass, $M_{H_D} = \mu_H-3\lambda V +\frac{(9c_1+30c_2)}{M_P}V^2$ to vanish to obtain the 2nd line of Eq.~(\ref{eq:mhc}).

The gauge coupling matching conditions are altered by the mass splitting of the octet and triplet, and become
\begin{align}
 \frac{1}{g_1^2(Q)}&=\frac{1}{g_5^2(Q)}+\frac{1}{8\pi^2}\biggl[
\frac{2}{5}
\ln \frac{Q}{M_{H_C}}-10\ln\frac{Q}{M_X}
\biggr]+\frac{8cV}{M_P} (-1)
~, \label{eq:matchg1} \\
 \frac{1}{g_2^2(Q)}&=\frac{1}{g_5^2(Q)}+\frac{1}{8\pi^2}\biggl[
2\ln \frac{Q}{M_{\Sigma_3}}-6\ln\frac{Q}{M_X}
\biggr]+\frac{8cV}{M_P} (-3)
~, \\
 \frac{1}{g_3^2(Q)}&=\frac{1}{g_5^2(Q)}+\frac{1}{8\pi^2}\biggl[
\ln \frac{Q}{M_{H_C}}+3\ln \frac{Q}{M_{\Sigma_8}}-4\ln\frac{Q}{M_X}
\biggr]+\frac{8cV}{M_P} (2)~.
\end{align}
From these equations, we can find
\begin{align}
 \frac{3}{g_2^2(Q)} - \frac{2}{g_3^2(Q)} -\frac{1}{g_1^2(Q)}
&=-\frac{3}{10\pi^2} \ln \left(\frac{QM_{\Sigma_3}^\frac{5}{2}}{M_{H_C}M_{\Sigma_8}^\frac{5}{2}}\right) 
-\frac{96cV}{M_P}
~,\label{eq:matchmhc} \\[3pt]
 \frac{5}{g_1^2(Q)} -\frac{3}{g_2^2(Q)} -\frac{2}{g_3^2(Q)}
&= -\frac{3}{4\pi^2}\ln\left(\frac{Q^6}{M_X^4 M_{\Sigma_3}M_{\Sigma_8}}\right) ~,
\label{eq:matchmgut}
\\[3pt]
 \frac{5}{g_1^2(Q)} +\frac{3}{g_2^2(Q)} -\frac{2}{g_3^2(Q)}&= -\frac{15}{2\pi^2} \ln\left(\frac{Q}{M_X}\right) + \frac{6}{g_5^2(Q)} -\frac{144cV}{M_P} ~.\label{eq:matchg5}
\end{align}
In the limit $\kappa_i\to 0$ and $M_{\Sigma_3},M_{\Sigma_8} \to M_{\Sigma}$, we recover the previous matching conditions given in \cite{Ellis:2016tjc,Ellis:2019fwf,Ellis:2020mno}. 

In our computations, we determine the GUT scale, $Q=M_{\rm GUT}$, by solving the condition $g_1(Q)=g_2(Q)$. Eqs.~(\ref{eq:matchmhc}-\ref{eq:matchg5}) provide three conditions that constrain the GUT scale parameters. So long as $c\ne0$, the parameters $\lambda$, $\lambda'$, $\kappa_1$, and $c_1 $ can remain free parameters. The coupling $\kappa_2$ can be absorbed into a redefinition of $V$ in the effective theory and the couplings $c_2$ and $c_3$ can be seen as small corrections to electroweak parameters.   
As discussed in Section~\ref{sec:result_kappa1}, $c_1$ effectively shifts $\lambda \to \tilde{\lambda}=\lambda-\frac{c_1}{M_P}V$. 
Now, we can use Eq.~\eqref{eq:matchmgut} at $Q=M_{\rm GUT}$ to determine the combination $M_G\equiv (M_X^4M_{\Sigma_3}M_{\Sigma_8})^{1/6}$.
By definition, $M_G$ is a function of the GUT breaking vev $V$ and the couplings ($g_5$, $\lambda$, $\lambda'$, $\kappa_1$). This then allows us to determine $V$ in terms of $M_G$ and the couplings ($g_5$, $\lambda$, $\lambda'$, $\kappa_1$). Once the vev is determined, it can then be substituted back into Eq.~(\ref{eq:matchg5}) to determine $g_5$, with ($\lambda$, $\lambda'$, $\kappa_1$) taken as inputs. Once $g_5$ is determined, Eq.~(\ref{eq:matchmhc}) can be used to determine $c$ and the constraints due to these matching conditions are satisfied.  

Next, we give the corrections to the gaugino masses~\cite{Hisano:1993zu,Evans:2019oyw,Tobe:2003yj},
\begin{align}
 M_1 &= \frac{g_1^2}{g_5^2} M_5
-\frac{g_1^2}{16\pi^2}\left[10 M_5 -10{A_\lambda}
 +\frac{48}{5}B_H\right]
-\frac{4cg_1^2v_0({A_\lambda}-B_H)}{M_P} ~,
\label{eq:m1match}
\\[3pt]
M_2 &= \frac{g_2^2}{g_5^2} M_5
-\frac{g_2^2}{16\pi^2}\left[6 M_5 - 6A_\lambda+6B_H -2B_\Sigma
 \right]
-\frac{12cg_2^2v_0(A_\lambda-B_H)}{M_P} ~,
\label{eq:m2match}
\\[3pt]
M_3 &= \frac{g_3^2}{g_5^2} M_5
-\frac{g_3^2}{16\pi^2}\left[4 M_5 - 4{A_\lambda}+3B_H -3B_\Sigma \right]
+\frac{8cg_3^2v_0(A_\lambda-B_H)}{M_P}
~,
\label{eq:m3match}
\end{align}
where we have used Eq.~\eqref{eq:DelT}, setting $\Delta_T \approx 0$. Here we can see that due to the tuning of the $B$ term, the gaugino matching conditions are quite similar to the case without the higher-dimensional operators.

The MSSM fields are embedded in the SU(5) multiplets as 
\begin{equation}
    \Psi \ni \{ Q_i,\, \bar{U}_i,\, \bar{E}_i \}~, \quad 
    \Phi \ni \{ \bar{D}_i ,\, L_i \} ~,
    \label{eq:embed}
\end{equation}
which gives the following MSSM Yukawa terms \footnote{For the superpotential which includes flavor mixing, see \cite{Ellis:2019fwf}.}
\begin{align}
 W_{\rm Yukawa}&=f_{u_{ij}}
(Q^{a}_i\cdot H_u)\overline{U}_{ja}-f_{d_{ij}} (Q^{a}_i\cdot H_d)
\overline{D}_{ja}-f_{e_{ij}}
\overline{E}_i (L_j\cdot H_d)\nonumber \\[2pt]
&-\frac{1}{2}f_{Q_iQ_{j}}
(Q^{a}_i \cdot Q^{b}_j) H^c_C
+f_{Q_iL_j}(Q^{a}_i\cdot L_j)\overline{H}_{Ca}
\nonumber \\[2pt]
&+f_{U_iE_{j}}\overline{U}_{ia}
\overline{E}_jH^a_C
-f_{U_iD_{j}}\epsilon^{abc}
\overline{U}_{ia}\overline{D}_{jb}\overline{H}_{Cc}~.
\label{eq:wyukawa}
\end{align}
These Yukawa couplings, including the higher-dimensional operator contributions, are defined as
\begin{eqnarray}
f_{u_{ij}}&=& 4\left(h_{\bf 10}\right)_{ij}- 3 c_{\Delta h,4}^{ij,S} R+\frac{1}{4}\left[3c_{\Delta h,3}^{ij,S}+5c_{\Delta h,3}^{ij,A}\right]R~,\nonumber \\
f_{Q_iQ_j}&=& 4\left(h_{\bf 10}\right)_{ij}+4 c_{\Delta h,4}^{ij,S}R-\frac{1}{4}c_{\Delta h,3}^{ij,S}R~,\label{eq:YukUp}\\
f_{U_iE_j}&=& 4\left(h_{\bf 10}\right)_{ij}+2 c_{\Delta h,4}^{ij,S}R-\frac{1}{2}\left[c_{\Delta h,3}^{ij,S}+5c_{\Delta h,3}^{ij,A}\right]R~, \nonumber 
\end{eqnarray}
\begin{eqnarray}
\sqrt{2} f_{d_{ij}}&=&\left(h_{\overline{\bf 5}}\right)_{ij}-3c_{\Delta h, 1}^{ij}R+2c_{\Delta h, 2}^{ij}R~,\nonumber \\
\sqrt{2} f_{e_{ij}}&=& \left(h_{\overline{\bf 5}}\right)_{ij}-3c_{\Delta h, 1}^{ij}R-3c_{\Delta h, 2}^{ij}R~, \label{eq:YukDown}\\
\sqrt{2} f_{Q_iL_j}&=& \left(h_{\overline{\bf 5}}\right)_{ij}+2c_{\Delta h, 1}^{ij}R-3c_{\Delta h, 2}^{ij}R~,\nonumber \\
\sqrt{2} f_{U_iD_j}&=& \left(h_{\overline{\bf 5}}\right)_{ij}+2c_{\Delta h, 1}^{ij}R+2c_{\Delta h, 2}^{ij}R~,\nonumber 
\end{eqnarray}
where $c^{ij,S}_{\Delta h,3} ~,\, c^{ij,S}_{\Delta h,4}$ denote the symmetric components and $c^{ij,A}_{\Delta h,3} $ denote the anti-symmetric components. We also note that $\left(h_{\bf 10}\right)_{ij}$ and $c^{ij}_{\Delta h,4}=c^{ij,S}_{\Delta h,4}$ are symmetric.

As can be seen in Eq.~(\ref{eq:YukUp}) and (\ref{eq:YukDown}), there are additional degrees of freedom beyond those needed to get the Yukawa coupling matching conditions. These additional degrees of freedom allow us to modify the coupling of the Standard Model fields to the colored Higgs. As was assumed in previous work\cite{Ellis:2019fwf}, we can use this freedom to take these couplings to be down quark- or lepton-like \cite{Ellis:2016tjc}. A more meaningful check on these couplings that arises when the higher-dimensional operators are considered is that the Planck-suppressed operators do not have coefficients larger than about one. Even with this restriction, the additional freedom provided by these higher-dimensional operators can be used to suppress significantly proton decay. 

We now discuss constraints on the $c_{\Delta h, k}$ parameters.  First, if we calculate the difference between the equations for $f_d$ and $f_e$ in Eq.~(\ref{eq:YukDown}), we find that it is determined by a single coupling.
\begin{eqnarray}
\sqrt{2}\left(f_{d_{ij}}-f_{e_{ij}}\right)=5c_{\Delta h, 2}^{ij}R ~.
\label{eq:c2}
\end{eqnarray}
Thus, the disparity between the lepton and down quark Yukawa couplings uniquely determines this coupling.
\SH{The matching of Eq.~\eqref{eq:c2} is determined at the GUT scale. Supersymmetry threshold corrections also contribute to the lepton and down quark Yukawa couplings, and the values of $f_{d_{ij}}$ and $f_{e_{ij}}$ depend on the supersymmetry parameters.
Besides, the value of $c_{\Delta h, 2}^{ij}$ also depends slightly on the supersymmetry parameter space, through the determination of the vev, hence $R$, and varies from roughly 0.0007 for the 1st generation to $\simeq 0.5$ for the 3rd generation, as discussed in more detail in Section \ref{depend}.}
The remaining freedom provided by the other higher-dimensional operators with couplings $c_{\Delta h, k}^{ij}$ only allows us to suppress one of the couplings $f_{QL}$ and $f_{UD}$, but not both. If we take $c_{\Delta h,3}^A=0$ in the up quark sector, then $f_u$, $f_{QQ}$, and $f_{UE}$ are determined by linearly-independent combinations of $h_{\bf 10}$, $c_{\Delta h,3}^S$ and $c_{\Delta h,4}$, and so are free parameters. This leaves significant room for additional suppression of the proton decay width beyond the down-like or lepton-like ansatz previously made \cite{Ellis:2016tjc}. This is relevant since it even allows for $p\to \pi^0 e^+$ or $p\to \pi^0 \mu^+$ to be the dominant decay mode.  These possibilities can be realized because the constraints on dimension-5 proton decay, which lead to a constraint on the SU(5) breaking vev, are weakened, allowing for a relative enhancement of the dimension-6 proton decay channels. In this case, dimension-6 proton decay channels may be close to the reach of Hyper-Kamiokande.

\section{Proton Decay}
\label{sec:pdecay}

\subsection{Colored Higgs Yukawa Couplings}

As discussed in the previous Section, including the contributions of Planck-suppressed operators  to the Yukawa sector may significantly change the proton decay predictions. In this Section, we examine these effects in detail. 

In the absence of the higher-dimensional operators, the Yukawa couplings $h_{10}$ and $h_{\bar 5}$ can be rotated to a basis where 
\begin{eqnarray}
&&h_{10} =\delta^{ij} e^{i\varphi_i}\hat{h}_{10}^i ~,\label{eq:h10Diag}\\
&& h_{\bar 5} = V_{ij}^*\hat h_{\bar 5}~, \label{eq:h5Diag}
\end{eqnarray}
where repeated indices are not summed, and $\sum_i \varphi_i = 0$~\cite{Ellis:1979hy}. In this basis, all redundant degrees of freedom have been removed. After GUT breaking, the fields are defined as   
\begin{equation}
    \Psi \ni \{ Q_i,\, e^{-i\varphi_i}\bar{U}_i,\, V_{ij}\bar{E}_j \}~, \quad 
    \Phi \ni \{ \bar{D}_i ,\, L_i \} ~,
    \label{eq:embed_diag}
\end{equation}
which reduces to the interaction basis for the Standard Model with no redundant parameters, and $V_{ij}$ can be identified as the Cabibbo-Kobayashi-Maskawa matrix. 

When the higher-dimensional operators are introduced, they explicitly break the GUT relations and each Standard Model fermion Yukawa coupling becomes independent below the GUT scale. Thus, instead of removing the redundant parameters from $h_{10}$ and $h_{\bar 5}$, the redundant parameters of $f_{u_{ij}}$, $f_{d_{ij}}$ and $f_{e_{ij}}$ need to be removed separately. This can be done easily.  However, if the couplings $c_{\Delta h,i}$ remain generic, determining the effect of these couplings on proton decay becomes an intractable challenge.  Since we only wish to discuss the general impact of these couplings on proton decay, we will take the simplifying ansatz\footnote{This ansatz is consistent with the flavor structure  of the Yukawa couplings necessary for suppressing the contribution from the Planck suppressed operators.}
\begin{eqnarray}
&&    c_{\Delta h,1} =V_{ij}^*  ( c_{\Delta h,1})_i ~, \label{eq:cdh1}\\
&&    c_{\Delta h,2} =V_{ij}^* (c_{\Delta h,2})_i ~, \label{eq:cdh2}\\
&&    c_{\Delta h,3} = \delta^{ij} e^{i\varphi_i}(c_{\Delta h,3})_i ~, \label{eq:cdh3}\\
&&    c_{\Delta h,4} =\delta^{ij} e^{i\varphi_i} (c_{\Delta h,4})_i ~, \label{eq:cdh4}
\end{eqnarray}
in the basis where the GUT Yukawa couplings are taken to be Eq.~(\ref{eq:h10Diag}) and (\ref{eq:h5Diag}), with $(c_{\Delta h,1-4})_i$ real. These couplings are chosen
so that the higher-dimensional operators have the same flavor structure as the Standard Model Yukawa couplings for the basis chosen in Eqs.~\eqref{eq:h10Diag} and \eqref{eq:h5Diag}.

Under these assumptions, we reduce the number of degrees of freedom from 66 to 9,
with the diagonal part of the colored Higgs Yukawa couplings taking the following form,
\begin{align}
f_{QQ,i} &= f_{u,i} +7(c_{\Delta h,4})_iR-(c_{\Delta h,3})_iR~,\nonumber\\
f_{UE,i} &= f_{u,i} +5(c_{\Delta h,4})_iR-\frac{5}{4}(c_{\Delta h,3})_i R~,
\label{eq:fqqfue}
\end{align}
where we have taken $c^A_{\Delta h,3} = 0$ and 
used the first line of Eq.~(\ref{eq:YukUp}) to substitute for $h_{\bf 10}$ in the expressions for $f_{QQ,i}$ and $f_{UE,i}$. 
Similarly, we use the first two lines of Eq.~(\ref{eq:YukDown}) to remove $h_{\overline{\bf 5}}$ and obtain 
\begin{align}
f_{QL,i} &= f_{e,i}+\frac{5}{\sqrt{2}}(c_{\Delta h, 1})_iR~,\nonumber \\
f_{UD,i} &= f_{d,i}+\frac{5}{\sqrt{2}}(c_{\Delta h, 1})_iR~.
\label{eq:fqlfud}
\end{align}
\SH{Eqs.~\eqref{eq:fqqfue} and \eqref{eq:fqlfud} determine the matching of colored Higgs Yukawa couplings and MSSM Yukawa couplings at the GUT scale, and then}
it is seen that $f_{QQ,i}$ and $f_{UE,i}$ are effectively free parameters that can be tuned by $(c_{\Delta h,3})_i$ and $(c_{\Delta h,4})_i$. This allows the couplings $f_{QQ,i}$ and $f_{UE,i}$ to be smaller than the Standard Model value of $f_{u,i}$. Furthermore, using $(c_{\Delta h, 1})_i$ we can also tune $f_{QL,i}$ or $f_{UD,i}$ to be significantly smaller than $f_{e,i}$ or $f_{d,i}$, respectively. 
\SH{In a supersymmetric model, due to the non-renormalization theorem~\cite{Weinberg:1998uv}, only wave function renormalizations contribute. Once the tuning is imposed at the GUT scale, the smallness of the coupling is protected through the RGEs.}
Thus, the Planck-suppressed operators can reduce the colored Higgs couplings to the Standard Model fields and reduce their contributions to baryon decay.

\subsection{Estimation}

To illustrate the possible impact of the Yukawa-type Planck-suppressed operators, we begin with a simplified and naive estimate that neglects renormalization-group (RG) running effects, in order to demonstrate how the proton decay modes can be modified. A more precise numerical analysis is presented in Section \ref{sec:results}. 
The dimension-five effective operators for proton decay can be written as
\begin{equation}
  {\cal L}_{5}^{\rm eff}
 =C_{5L}^{i j k l}{\cal O}^{5L}_{i j k l}
 +C_{5R}^{i j k l}{\cal O}^{5R}_{i j k l} + {\rm h.c.}
 ~,
 \label{eq:l5eff}
 \end{equation}
 with 
 \begin{align}
  {\cal O}^{5L}_{i j k l}&=\int d^2\theta ~
 \frac{1}{2}\epsilon_{abc}\epsilon_{mn}\epsilon_{pq}~
 Q^{am}_iQ^{bn}_j
 Q_k^{cp}L^q_l~,
  \nonumber\\
 {\cal O}^{5R}_{i j k l}&=\int d^2\theta ~
 \epsilon^{abc}
  \bigl(\bar{U}_i\bigr)_a
  \bar{E} _j
 \bigl(\bar{U} _k\bigr)_b
 \bigl(\bar{D} _l\bigr)_c ~,
\end{align}
where $i,j,\dots$ denote the generation indices, $a,b,c$ the color indices, and $p,q$ the $\mathrm{SU}(2)_L$ indices; $\epsilon_{abc}$ and $\epsilon_{pq}$ are the Levi-Civita symbols. 
If we assume that the $c_{\Delta h,i}$ have a similar flavor structure to the Yukawa couplings, as in Eqs.~(\ref{eq:cdh1})-(\ref{eq:cdh4}), their effects on the flavor diagonalization can be neglected, as stated previously. The Wilson coefficients at the GUT scale are given by \cite{Ellis:2016tjc}
\begin{align}
  C^{i j k l}_{5L}  (M_{\rm GUT}) &= \frac{1}{M_{H_C}} f_{QQ,i}e^{i\varphi_i}\delta^{ij}(V^*)^{kl}f_{QL,l} ~,
\nonumber \\
  C^{i j k l}_{5R} (M_{\rm GUT}) & = \frac{1}{M_{H_C}} f_{UE,i} V^{ij}(V^*)^{kl}f_{UD,l}e^{-i\varphi_k} ~.
 \label{eq:dim5gutmatchijkl}
\end{align}
If we assume that the $c_{\Delta h,i}$ have similar flavor structure to the Yukawa couplings, only operators $ {\cal O}^{5L}_{i i 1 j}$ and $ {\cal O}^{5R}_{331k}$ with $i=2,3$, $j=1,2,3$ and $k=1,2$ give sizable contributions~\footnote{Our primary interest is the potential suppression of proton decays induced by Planck-suppressed operators. We therefore omit discussion of the Wilson-coefficient components that already yield negligible contributions prior to the inclusion of the Planck-suppressed operators.
}.

At the supersymmetry-breaking scale $M_{\rm SUSY}$, sfermions are integrated out via Wino- or Higgsino-exchange processes at one-loop order, and these coefficients are matched onto the coefficients of the following effective operators \cite{Ellis:2016tjc}
\begin{align}
  {\cal L}^{\text{eff}}_{\rm SM}&=C^{\widetilde{H}}_i {\cal O}_{1i33}
 + C^{\widetilde{W}}_{jk}\widetilde{\cal O}_{1jjk}
 + C^{\widetilde{W}}_{jk}\widetilde{\cal O}_{j1jk}
 + \bar{C}^{\widetilde{W}}_{jk}\widetilde{\cal O}_{jj1k}
 ~,
\end{align}
with $i =1,2$, $j=2,3$, and $k=1,2,3$,
where the effective operators have the form
\begin{align}
 {\cal O}_{ijkl} &\equiv \epsilon_{abc}(u^a_{Ri}d^b_{Rj})
(Q_{Lk}^c \cdot L_{Ll}) ~, \nonumber \\
 \widetilde{\cal O}_{ijkl} &\equiv \epsilon_{abc} \epsilon^{\alpha\beta}
\epsilon^{\gamma\delta} (Q^a_{Li\alpha}Q^b_{Lj\gamma})
(Q_{Lk\delta}^c L_{Ll\beta}) ~.
\label{eq:effopPD}
\end{align}
We have 
\begin{align}
    \label{eq:CH}
	C_i^{\widetilde{H}} (M_{\rm SUSY}) &=\frac{f_tf_\tau}{(4\pi)^2} C^{*331i}_{5R}(M_{\rm SUSY}) F(\mu, m_{\widetilde{t}_R}^2, m_{\widetilde{\tau}_R}^2) ~, \\ 
	C_{jk}^{\widetilde{W}} (M_{\rm SUSY}) &=\frac{\alpha_2}{4\pi} C^{jj1k}_{5L}(M_{\rm SUSY}) \bigl[ F(M_2, m_{\widetilde{Q}_1}^2, m_{\widetilde{Q}_j}^2) +F(M_2, m_{\widetilde{Q}_j}^2, m_{\widetilde{L}_k}^2) \bigr] ~,\nonumber \\ 
	\bar{C}_{jk}^{\widetilde{W}} (M_{\rm SUSY}) &= -\frac{3}{2}\frac{\alpha_2}{4\pi} C^{jj1k}_{5L}(M_{\rm SUSY}) \bigl[ F(M_2, m_{\widetilde{Q}_j}^2, m_{\widetilde{Q}_j}^2) +F(M_2, m_{\widetilde{Q}_1}^2, m_{\widetilde{L}_k}^2) \bigr] ~,
    \label{eq:CW}
\end{align}
where
\begin{equation}
	F(M, m_1^2, m_2^2) \equiv \frac{M}{m_1^2-m_2^2} \biggl[ \frac{m_1^2}{m_1^2-M^2}\ln \biggl( \frac{m_1^2}{M^2}\biggr) - \frac{m_2^2}{m_2^2-M^2}\ln \biggl( \frac{m_2^2}{M^2}\biggr) \biggr] ~. \label{eq:F}
\end{equation}
The coefficients in $C_i^{\widetilde{H}}$ and $C_{jk}^{\widetilde{W}},\bar{C}_{jk}^{\widetilde{W}}$  that can give the most important contributions to proton decay are
\begin{align}
	 C^{331i}_{5R} & = \frac{1}{M_{H_C}} (f_{UE})_3 V^{33}(V^*)^{1i}(f_{UD})_{i=1,2}e^{-i\varphi_1} ~,\nonumber \\ 	 C^{jj1k}_{5L} & =\frac{1}{M_{H_C}} (f_{QQ})_{j=2,3}e^{i\varphi_j}(V^*)^{1k}(f_{QL})_{k=1,2,3} ~.
    \label{eq:CHCW}
\end{align}
As seen in Eq.~\eqref{eq:fqqfue}, all of the relevant Wilson coefficients can be suppressed. However, at least one of $(f_{QQ})_3$ and $(f_{UE})_3$ has to be tuned, which means $(c_{\Delta h,3})_{3}$ and $(c_{\Delta h,4})_{3}$ need to be large (at least for the third generation). Values for these couplings will be discussed further in Section \ref{depend}. 
In most cases, the contribution from the Wino-exchange process is much larger than the contribution from the Higgsino-exchange process. Thus, we can choose to suppress $f_{QL,i}$ in Eq.~\eqref{eq:fqlfud} and leave $f_{UD,i}$ to be determined by $f_{QL,i}$.

\SH{As shown above, the quantities $(c_{\Delta h,1})_i$, $(c_{\Delta h,3})_i$ and $(c_{\Delta h,4})_i$ that parametrize Planck-suppressed operators can suppress dimension-5 proton decay. Dimension-6 proton decay is mediated by gauge boson exchange, and its contribution to the Wilson coefficients is determined by the gauge boson mass $M_X$ and the matter content embedding. This is not affected by $(c_{\Delta h,1})_i$, $(c_{\Delta h,3})_i$ and $(c_{\Delta h,4})_i$ in our ansatz.
As will be shown in Section~\ref{sec:results}, dimension-6 proton decay is beyond the reach of current experimental bounds. 
However, if we considered the more general setup $(c_{\Delta h, 1})_{ij}$, $(c_{\Delta h, 2})_{ij}$, $(c_{\Delta h, 3})_{ij}$, and $(c_{\Delta h, 4})_{ij}$ and include the off-diagonal components, these may change the mass diagonalization and therefore affect the dimension-6 proton decay widths.
}

\subsection{Nucleon decay modes}

We consider the nucleon decay modes
$p \to \pi^0 e^+, \pi^0 \mu^+, K^+ \bar \nu, \pi^+ \bar \nu$, $K^0 e^+, K^0 \mu^+$ and $n\to \pi^0 \bar \nu$, $\pi^- e^+$, $\pi^- \mu^+, K^0 \bar \nu$.
Neutron decay widths can be related to proton decay widths via SU(2) isospin relations as
\begin{align}
    &\Gamma (n \to \pi^0\bar{\nu}) = \frac{1}{2} \Gamma(p\to \pi^+\bar{\nu}) ~,\nonumber\\
    &\Gamma (n \to \pi^-e^+) = 2\Gamma(p \to \pi^0 e^+) ~,\nonumber\\
    &\Gamma (n \to \pi^-\mu^+) = 2\Gamma(p \to \pi^0 \mu^+) ~,\nonumber\\
    &\Gamma (n \to K^0\bar{\nu}) = \Gamma(p \to K^+\bar{\nu}) ~.
\end{align}
We retain the mode $n\to \pi^0 \bar \nu$ because it has higher experimental sensitivity than that for $p\to \pi^+\bar{\nu}$, as shown in Table~\ref{tab:channels} where we display the current limits on the nucleon lifetime for each channel considered \cite{Super-Kamiokande:2014otb, Takhistov:2016eqm,Super-Kamiokande:2025lxa,Super-Kamiokande:2020wjk,PhysRevD.72.052007,Super-Kamiokande:2022egr} as well as some projecected limits from Hyper-Kamiokande \cite{Hyper-Kamiokande:2018ofw}.~\footnote{The sensitivities achieved by JUNO and DUNE are expected to be inferior to Hyper-Kamiokande sensitivities. For example, the JUNO sensitivity for $p\to K^+\bar{\nu}$ is expected to be $9.6 \times 10^{33}$~years~\cite{JUNO:2022qgr,JUNO:2021vlw}.} 

\begin{table} [ht]
  \centering
  \begin{tabular}{lll}
  \hline \hline
   Decay Mode ~~ & Current  [years] ~~ & HK sensitivity  [years]  \\
  \hline
   $p \to K^+ \bar{\nu} $ & $6.6 \times 10^{33}$~\cite{Super-Kamiokande:2014otb, Takhistov:2016eqm} & $3.2 \times 10^{34}$~\cite{Hyper-Kamiokande:2018ofw} \\
   $p \to \pi^+ \bar{\nu} $ & $3.5 \times 10^{32}$~\cite{Super-Kamiokande:2025lxa} & \\
   $n \to \pi^0 \bar{\nu}$ & $1.4 \times 10^{33}$~\cite{Super-Kamiokande:2025lxa}  & \\
   $p \to \pi^0 e^+ $ & $2.4 \times 10^{34}$~\cite{Super-Kamiokande:2020wjk} & $7.8 \times 10^{34}$~\cite{Hyper-Kamiokande:2018ofw}  \\
   $p \to \pi^0 \mu^+ $ & $1.6 \times 10^{34}$~\cite{Super-Kamiokande:2020wjk} & $7.7 \times 10^{34}$~\cite{Hyper-Kamiokande:2018ofw} \\ 
   $p \to K^0 e^+ $ & $ 1.0\times 10^{33}$~\cite{PhysRevD.72.052007} &  \\ 
   $p \to K^0 \mu^+ $ & $ 3.6\times 10^{33}$~\cite{Super-Kamiokande:2022egr} &  \\ 
  \hline \hline
  \end{tabular}
  \caption{Current 90\% CL limits on nucleon decay modes from Super-Kamiokande and projected 90\% CL sensitivities for a 1.9~Megaton$\cdot$year exposure of Hyper-Kamiokande.}
  \label{tab:channels}
\end{table}

As can be seen from Eq.~\eqref{eq:dim5gutmatchijkl}, dimension-5 proton decay modes are suppressed by the Yukawa couplings, while dimension-6 decay modes are not. Neutrino modes that include the third-generation particles tend to provide larger contributions to the dimension-5 proton partial decay widths. As for $e^+$ and $\mu^+$ final states, the Yukawa couplings in the dimension-5 Wilson coefficients will give ${\cal O}(10^{-4}-10^{-8})$ suppression compared to the corresponding neutrino modes, so the dimension-6 proton decay modes give a larger contribution.
Hence $p \to \pi^0 e^+, \pi^0 \mu^+, K^0 e^+, K^0 \mu^+$ are dimension-6 dominated decay modes while $p\to K^+ \bar \nu, \pi^+ \bar \nu$ and $n\to \pi^0 \bar \nu$ are dimension-5 dominated decay modes.
In our ansatz, where the fields can be expressed as in Eq.~\eqref{eq:embed_diag} after the GUT breaking, the Planck-suppressed operators contributing to the Yukawa couplings do not affect the dimension-6 proton decay modes. Planck-suppressed operators in the Higgs sector can change the matching conditions and hence the GUT mass spectrum. However, its effect on the dimension-6 and dimension-5 proton decay widths will be an overall factor.

\section{Results}
\label{sec:results}

In this Section, we present numerical calculations of nucleon decay rates when the effects of various dimension-5 Planck-suppressed operators are included.
Previous work\cite{Ellis:2016tjc,Ellis:2019fwf} on this subject primarily focused on the effect of the higher-dimensional operator proportional to $c$. It was found that $c$ could be used to satisfy the gauge coupling matching conditions while leaving $\lambda$ and $\lambda'$ free parameters. With $\lambda'$ a free parameter, the SU(5) breaking vev could be larger, which then increased the predicted proton lifetime.

Here, we examine the effects of other higher-dimensional operators that can enhance the proton's lifetime. First, we will examine a scenario where $\lambda'$ is sufficiently small that the higher-dimensional operators proportional to $\kappa_1$ and $\kappa_2$ set the size of the SU(5) breaking vev instead of $\lambda'$. This actually has an important effect on the proton decay width due to the fact that the masses of $\Sigma_3$ and $\Sigma_8$ are now split. In this scenario, the higher-dimensional operators proportional to $c$, $\kappa_1$, and $\kappa_2$ are included. They alter the gauge coupling matching conditions in Eq. (\ref{eq:matchmhc}). This allows for a larger colored Higgs boson mass, which then suppresses the proton decay width. Here $c$ plays the same role as it did in \cite{Ellis:2016tjc,Ellis:2019fwf}, i.e. allows us to take $\kappa_1$ and $\kappa_2$ as free parameters.

After examining this different breaking pattern for SU(5), we examine the effects of the higher-dimensional operators affecting the Yukawa couplings of the colored Higgs bosons. We will see that including these operators allow us to tune the couplings of the colored Higgs boson leading to a significant deviation from the standard proton decay signatures of minimal SU(5). When these types of scenarios are considered, we will assume that $\kappa_1$ and $\kappa_2$ are small enough that the SU(5) breaking vev is set predominantly by $\lambda'$ and $m_{\Sigma_3}=m_{\Sigma_8}$.

In the majority of what follows, we consider a CMSSM-like parameter space\footnote{Despite the present lack of experimental signals, regions of the CMSSM parameter space still survive \cite{Ellis:2022emx,Antusch:2025rbp}.}. For each scenario, we take 
\begin{equation}
 m_0,\ m_{1/2},\ A_0,\ \lambda,\ \lambda',\ \tan \beta,\ {\rm
  sign}(\mu) \, ,\label{eq:FreeParamAll}
\end{equation}
as free parameters. We assume here that the universality scale (the scale at which all scalar and gaugino masses and $A$-terms are unified) is the same as the GUT scale defined to be where $g_1 = g_2$.  These parameters will be supplemented by other parameters depending on the scenario considered. In particular, 
the following couplings associated with Planck-suppressed operators will play a role in at least some part of the results below, 
\begin{equation}
c,\  \kappa_1,\  c_{\Delta h, 1},\  c_{\Delta h, 2},\ c_{\Delta h, 3},\ c_{\Delta h, 4}\
\, .\label{eq:FreeParamPlanck}
\end{equation}
The coupling $c$ will be used throughout and is used to satisfy the gauge matching conditions. The couplings 
\begin{equation}
c_1,\ c_2,\ c_3,\ \kappa_2\  
\, ,\label{eq:FreeParamAbs}
\end{equation}
will have a subleading effect compared to the operators we consider. The couplings $c_1$ and $c_2$ alter the $\mu$ parameter, but this effect is generically much smaller than the size of the Higgs vev. Since $\mu$ is generally of order a TeV and above, this is a very small effect. Furthermore, its effect can be absorbed into our definition of $\mu$ to leading order. $c_1$ also affects the colored Higgs boson mass. However, this contribution is smaller than the contribution from renormalizable operators, since we generically need $\lambda\sim 1$. This correction to the colored Higgs boson mass, thus, has no significant effect on the proton lifetime. The coupling $c_3$ leads to a correction to the Higgs potential suppressed by the Planck scale and is thus very insignificant. Since $\kappa_2$ does not affect $M_{\Sigma_3}$ and $M_{\Sigma_8}$, its only effect is to alter the vev that breaks SU(5). However, the SU(5) breaking vev is constrained by the gauge matching conditions negating any potential effect. Because of these constraints, $\kappa_2$ only leads to an altered relationship between $\mu_\Sigma$ and the SU(5) breaking scale. However, this has no effect on the proton lifetime that is calculated below.   
Finally, the couplings of the Planck-suppressed operators that we do not consider in this work (for reasons given above) are:
\begin{equation}
c_{\Sigma\Sigma},\ c_{\bar{H}\Sigma H},\ c_{F\Sigma},\ c_{L\!\!\!/},\ c_{B\!\!\!\!/L\!\!\!/}\ \, .\label{eq:FreeParamNU}
\end{equation}

Our computation follows the formalism outlined in Refs.~\cite{Hisano:2013exa, Nagata:2013sba, Nagata:2013ive, Evans:2015bxa, Ellis:2015rya}. 
We perform two-loop RGE running for the gauge couplings and Yukawa couplings, and one-loop running for the Wilson coefficients. The one-loop threshold corrections are also included for the gauge and Yukawa couplings. For the Yukawa sector discussion, we focus on the decay modes $p\to K^+ \bar \nu, \pi^+ \bar \nu$ and $n\to \pi^0 \bar \nu$, since colored Higgs Yukawa couplings do not affect dimension-6 proton decays directly.
Table~\ref{tab:channels} summarizes the current experimental limits on various proton decay modes, as well as projected limits from Hyper-Kamiokande.

\subsection{Proton Decay and SU(5) Breaking}
\label{sec:result_kappa1}

In this Section we focus on the Planck-suppressed operators that affect the SU(5) Higgs sector. The only two operators which have a significant effect on SU(5) breaking are $\kappa_1$ and $\kappa_2$. However, as discussed above, $\kappa_2$ will have no significant effect on the proton lifetime and so will be ignored here.  

We also keep the Planck-suppressed operator proportional to $c_{\Delta h,2}$, since it is needed to realize the splitting in the lepton and down quark Yukawa couplings. Without this coupling, the theory would be ill defined. Including this operator has no direct effect on the proton lifetime, since its effect is completely absorbed into the definition of $f_e$ and $f_d$. In addition to this, we define the colored Higgs boson Yukawa couplings in terms of these $c_{\Delta h, 2}$ adjusted Standard Model Yukawa couplings plus other Planck-suppressed operators, as seen in Eqs.~\eqref{eq:fqqfue} and \eqref{eq:fqlfud}. With this in mind, when we say no other Planck-suppressed operators are considered, we mean $f_{QL}=f_e$, $f_{UD}=f_d$, and $f_{QQ}=f_{UE}=f_u$. This is in contrast to what was done in previous work \cite{Ellis:2015rya,Ellis:2016tjc,Ellis:2017djk,Ellis:2019fwf,Ellis:2020mno,Evans:2021hyx}, where ignorance of the details of the Planck-suppressed operators was used to assume that $f_{QL}=f_d$ at the GUT scale. This will lead to an apparent suppression of the proton lifetime in this paper, even for the scenarios we say have no Planck-suppressed operators affecting the colored Higgs boson Yukawa couplings, as will be seen later in this Section. 

To determine the effect of $\kappa_1$ on the proton lifetime, we examine the matching conditions in Eq.~\eqref{eq:matchmhc}, \eqref{eq:matchmgut} and \eqref{eq:matchg5} for $\lambda'=0$ and $\kappa_1\ne 0$, with the $\Sigma_3$ and $\Sigma_8$ masses determined by Eq. (\ref{eq:Msig}). As mentioned above, we also retain the $c$ parameter to ensure that these matching conditions can be satisfied for a generic $\kappa_1$. 
To highlight the effect of $\kappa_1$, we will compare two limits: (i) $M_{\Sigma_3}\approx \frac{5}{2}\lambda'V~,\,\, M_{\Sigma_8}\approx \frac{5}{2}\lambda'V$ and, (ii) $M_{\Sigma_3}\approx 20\frac{\kappa_1V^2}{M_P}~,\,\, M_{\Sigma_8}\approx 5\frac{\kappa_1V^2}{M_P}$~. Limit (i) has been considered in previous work \cite{Ellis:2015rya,Ellis:2016tjc,Ellis:2017djk,Ellis:2019fwf,Ellis:2020mno,Evans:2021hyx}. Limit (ii), on the other hand, describes the case where the Planck-suppressed operators proportional to $\kappa_1$ and $\kappa_2$ determine the SU(5) breaking vev~\footnote{In limit (i), where renormalizable operators dominate, the couplings of the renormalizable operators must be of order $\lambda\sim 1$ and $\lambda'\sim 10^{-4}$ in order to get a sufficiently long proton lifetime. We regard limit (ii) as a plausible alternative.}.

In both these limits the matching condition in Eq.~\eqref{eq:matchmgut} still constrains the same set of GUT particle masses, which we define as $M_G$. However, the dependence of $M_G$ on the SU(5) breaking vev is very different in the two limits. In limit (i), $M_G$, which is completely determined by Eq.~\eqref{eq:matchmgut}, depends on $\lambda'$ and $g_5$:
\begin{align}
    M_G^6 &= M_X^4M_{\Sigma_3}M_{\Sigma_8} = g_5^4\left(\frac{\lambda'}{2}\right)^2(5V)^6 ~.\label{eq:MG1}
\end{align}
This then allows us to rewrite the colored Higgs boson mass as a function of $M_G$ instead of $V$, 
\begin{align}
    M_{H_C} &= \lambda \biggl( \frac{2}{\lambda'g_5^2}\biggr)^{\frac{1}{3}}M_G~.
    \label{eq:MG1HC}
\end{align}
If these expressions for $M_G$, $M_{H_C}$ are substituted back into the gauge matching conditions in Eq.~\eqref{eq:matchmhc}, Eq.~\eqref{eq:matchmgut}, and Eq.~\eqref{eq:matchg5}, we are left with three equations and three unknowns ($c$, $g_5$, $M_G$), which can be solved for.  

In the case of limit (ii), an identical procedure can be followed to determine the colored Higgs boson mass. However, the free parameter $\lambda'$ is replaced by $\kappa_1$ and the dependence on all these couplings is altered due to the fact that $M_{\Sigma_{3,8}}$ depend on $V$ quadratically, not just linearly. In this case, we can again derive the colored Higgs boson mass as a function of $M_G$. However, in this case the functional form is now:
\begin{align}
    M_G^6 &= M_X^4M_{\Sigma_3}M_{\Sigma_8} = 2^25^6g_5^4\left(\frac{\kappa_1}{M_P}\right)^2V^8 ~,\\
    M_{H_C} &= \lambda \biggl( \frac{5}{2g_5^2(\frac{\kappa_1}{M_P})} \biggr)^{\frac{1}{4}} M_G^{\frac{3}{4}}~.
    \label{eq:MG2}
\end{align}
The process to determine ($c$, $g_5$, and $\kappa_1$) is now the same as the previous case, with the exception that, since $M_{\Sigma_3}/M_{\Sigma_8}\simeq 4$, Eq.~\eqref{eq:matchmhc} has an additional factor of $4^{5/2}$.

As is clear from comparing these expressions, the relationship between the colored Higgs mass, which is important in determining the proton lifetime, and $M_G$ is altered. This, plus the fact that the matching condition in Eq. (\ref{eq:matchmhc}) is altered by the mass splitting of $\Sigma_3$ and $\Sigma_8$, leads to a different proton lifetime for a given set of parameters. 

Before discussing the main results of this Section, we point out how our calculation of the proton lifetime is different in this Section than was done in previous work \cite{Ellis:2015rya,Ellis:2016tjc,Ellis:2016qra,Ellis:2017djk,Ellis:2019fwf,Ellis:2020mno,Evans:2021hyx}. In these previous works, the colored Higgs Yukawa coupling was set equal to which ever was smaller of the down type quark Yukawa couplings of the lepton type Yukawa couplings. This choice was made based on the understanding that there were higher-dimensional operators contributing to these Yukawa couplings and they would lead to a contribution of order one of these couplings. In this work, we do a much more rigorous treatment of these Yukawa couplings. Here, when we take the contribution to the colored Higgs Yukawa couplings from higher-dimensional operators to be zero, we then must set these Yukawa couplings equal to the contributions found in Eqs.~(\ref{eq:fqqfue}) and (\ref{eq:fqlfud}) for $c_{\Delta h,i=1,3,4}=0$.  This necessarily leads to some Yukawa couplings being defined in terms of the down quark Yukawa couplings and others in terms of the lepton Yukawa couplings. This inevitably leads to a suppression in the proton lifetime prediction compared to previous work.

\begin{figure}[ht!]
    \centering
    \vspace{-2mm}
    \includegraphics[width=0.45\linewidth]{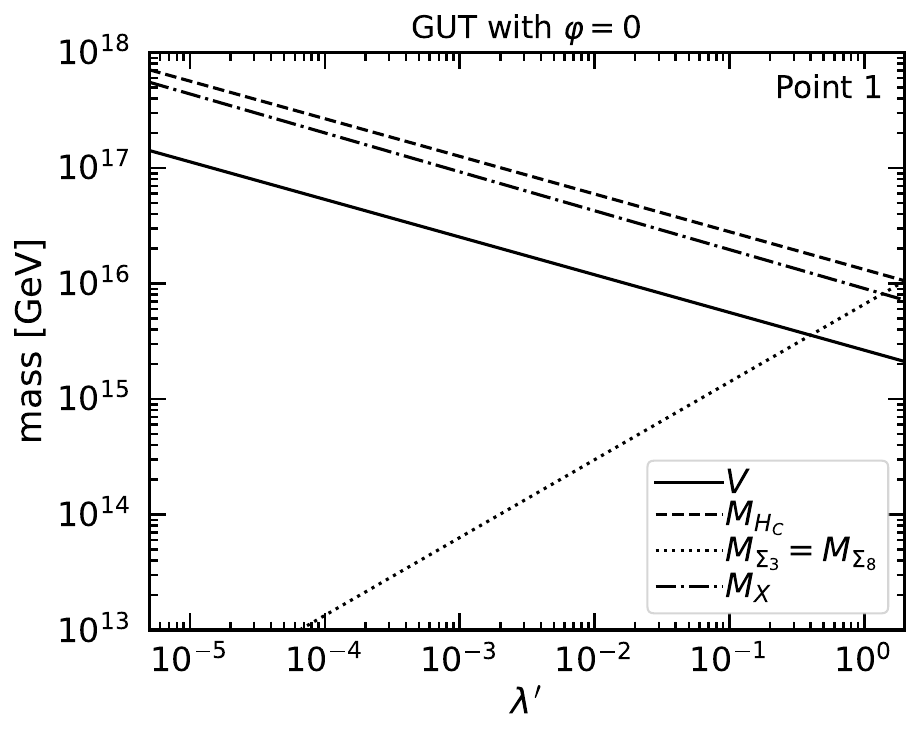}
    \vspace{-2mm}
    \includegraphics[width=0.45\linewidth]{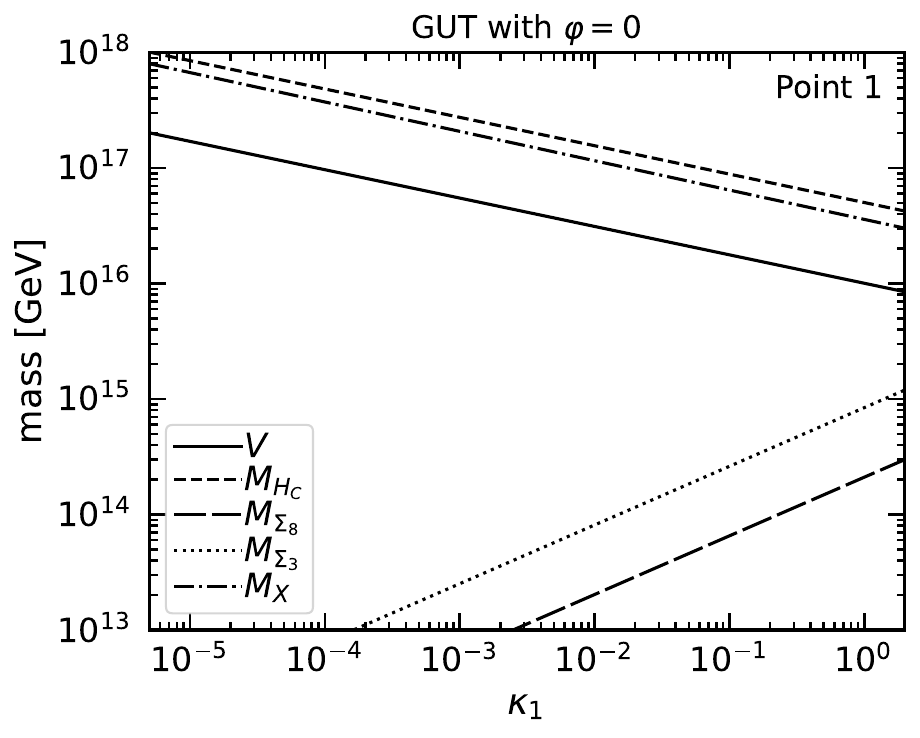}
    \vspace{-2mm}
    \includegraphics[width=0.45\linewidth]{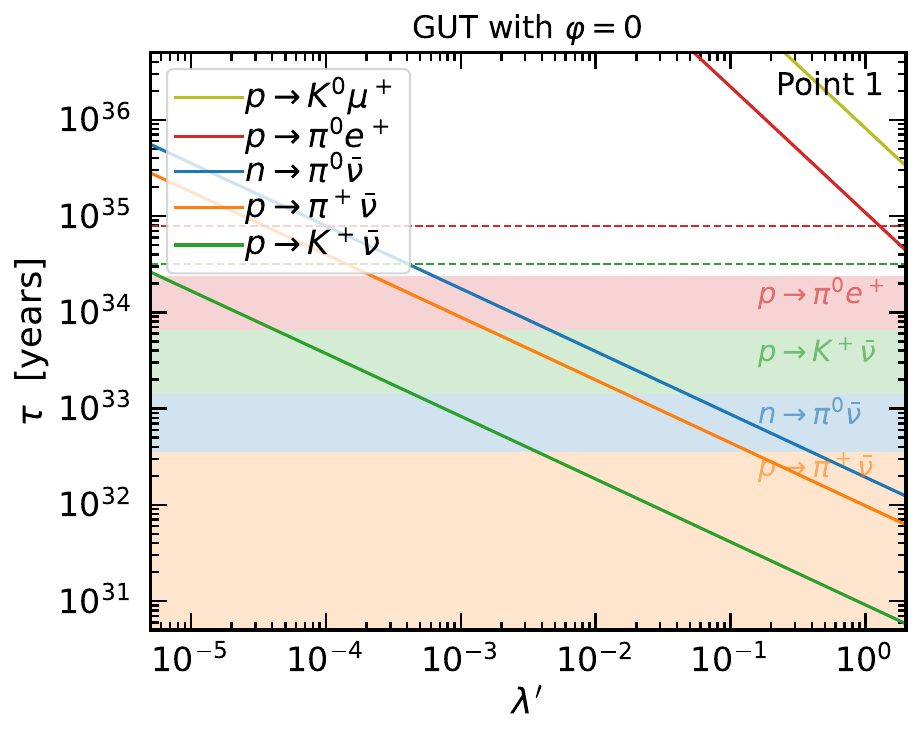}
    \vspace{-2mm}
    \includegraphics[width=0.45\linewidth]{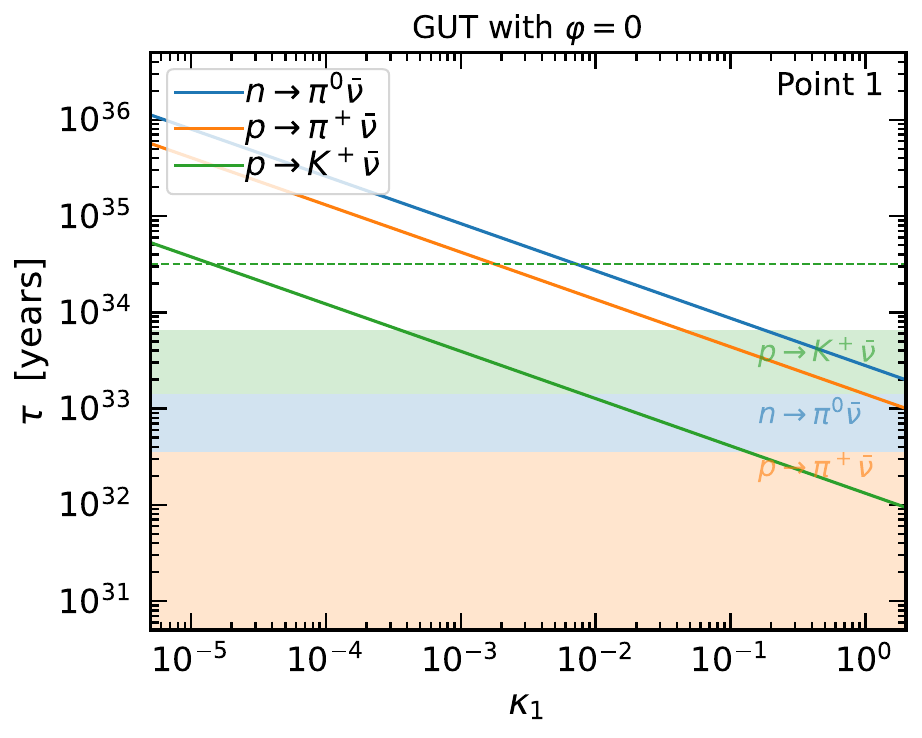}
    \vspace{-2mm}
    \caption{Left: GUT mass spectra and the dependences of nucleon lifetimes  on $\lambda'$ in limit (i), where we set $\lambda=1$, $\kappa_1=0$. Right: GUT mass spectra and the dependences of nucleon lifetimes on $\kappa_1$ in limit (ii), where we set $\lambda=1$, $\lambda'=0$. The GUT phases $\varphi_i$ are assumed to vanish in both scenarios, and the supersymmetry parameters taken to be those of Point 1 given in Table~\ref{tab:point1}. The color-coded solid lines in the lower panels are the calculated nucleon lifetimes, the shaded regions show the current constraints on the corresponding nucleon lifetimes, and the dashed lines show prospective Hyper-Kamiokande sensitivities.}  
    \label{fig:kappa}
\end{figure}

Fig.~\ref{fig:kappa} illustrates the dependence of the GUT mass spectra and proton lifetimes on $\lambda'$ in limit (i) and on $\kappa_1$ in limit (ii). The input parameters for each figure correspond to an example
that produces the correct dark matter  abundance and Higgs boson mass~\footnote{We choose points that are similar to those chosen in \cite{Ellis:2019fwf}, with slight variations to account for updates to experimentally measured couplings, etc. } 
\begin{itemize}
    \item Point 1: $m_0=27.9$~TeV, $m_{1/2}=9.5$~TeV, $A_0=0$, $\tan \beta=4$, $\mu>0$. $\lambda=\lambda'=1$~,
\end{itemize}
which we will refer to as Benchmark point 1. The inputs, derived GUT-scale quantities, MSSM parameters and observables for this point is summarized in Table~\ref{tab:point1}~\footnote{
We use {\tt FeynHiggs 2.18.1} to compute the Higgs mass~\cite{Heinemeyer:1998yj,Heinemeyer:1998np,Degrassi:2002fi,Frank:2006yh,Hahn:2013ria,Bahl:2016brp,Bahl:2017aev,Bahl:2018qog}. We have tested {\tt FeynHiggs 2.16.1, 2.17.0} and {\tt 2.18.1} and obtained results that are similar within ${\cal O}(1~{\rm GeV})$. }. As is seen in the figure (which drops the condition that $\lambda' = 1$),  both limits can enhance the proton lifetime.  However, the proton lifetimes and GUT mass spectra are different and lead to different coupling-dependent scaling.  

\begin{table}[ht]
  \centering
  \begin{tabular}{cccc}
  \hline \hline
  \multicolumn{4}{c}{Inputs}\\
  \hline
  $m_{1/2}=9.5$~TeV & $m_0=27.9$~TeV & $A_0/m_0=0$ & $\tan{\beta}=4$ \\
  $\lambda=1$ & $\lambda'=1$ & & \\
  \hline
  \multicolumn{4}{c}{GUT-scale parameters (masses in units of $10^{16}~$GeV)}\\
  \hline
  $M_{\rm GUT}=0.818$& $M_{H_C}=1.322$ & $M_\Sigma=0.661$ & $M_X=0.906$ \\
  $V=0.264$ & $g_5=0.685$ & $c=0.1365$ \\
  \hline
  \multicolumn{4}{c}{MSSM parameters (masses in units of TeV)}\\
  \hline
  $m_\chi=0.987$ & $m_{\tilde{t}_1}=21.3$ & $m_{\tilde{g}}=19.40$ & $m_{\chi_2}=0.988$ \\
  $m_{A}=28.85$ 
  & $\mu=0.938$ & $m_{\tilde{\ell}_L}=28.4$& $m_{\tilde{\ell}_R}=28.0$ \\
  $m_{\tilde{\tau}_1}=28.0$ & $m_{\tilde{q}_L}=31.2$ & $m_{\tilde{d}_R}=30.8$ & $m_{\tilde{t}_2}=27.0$ \\
  $A_t=18.45$ & $A_b=27.2$ & $B=-218$ & \\
  \hline
  \multicolumn{4}{c}{Observables}\\
  \hline
  $\Omega_\chi h^2=0.096$ & $m_h=122.9$~GeV & & \\
  \hline \hline
  \end{tabular}
  \caption{Benchmark point 1 parameters. The observables in the last line are the dark matter relic density and Higgs mass calculated with {\tt FeynHiggs 2.18.1}. 
  }
  \label{tab:point1}
\end{table}

We show in the upper panels of Fig.~\ref{fig:kappa} the dependence of the GUT mass spectrum on $\lambda'$ in limit (i), where we set $\lambda=1$, $\kappa_1=0$ (left plot). and the dependence of the GUT mass spectrum on $\kappa_1$ in limit (ii), where we set $\lambda=1$, $\lambda’=0$ (right plot). The color-coded lines in the lower plots show how nucleon decay lifetimes vary with $\lambda’$ and $\kappa_1$, respectively, reflecting the variations in the GUT masses. The regions shaded in corresponding colors show the current constraints on the nucleon lifetimes, and the dashed lines show prospective Hyper-Kamiokande sensitivities. We see in the lower left panel of Fig.~\ref{fig:kappa} that the strongest constraint in limit (i) comes from $p \to K^+ \bar \nu$ decay, which requires $\lambda’ \lesssim 5 \times 10^{-5}$ (corresponding to $M_X \gtrsim 3 \times 10^{17}$~GeV)~\footnote{\SH{Due to the non-renormalization theorem~\cite{Weinberg:1998uv} in supersymmetric models, the couplings $\lambda'$ and $\kappa_1$ do not receive radiative corrections, and the small sizes of these parameters are protected by the RGEs.}}. 
Searches for $n \to \pi^0 \bar \nu$ and $p \to \pi^+ \bar \nu$ impose weaker constraints, $\lambda’ \lesssim 10^{-1}$, and no useful constraints are imposed by $p \to \pi^0 e^+$ and $p \to K^0 \mu^+$ searches. We see in the lower right panel of Fig.~\ref{fig:kappa} that the strongest constraint in limit (ii) again comes from $p \to K^+ \bar \nu$ decay ($\kappa_1 \lesssim 4 \times 10^{-4}$), with no useful constraints from $n \to \pi^0 \bar \nu$ and $p \to \pi^+ \bar \nu$ searches.

From Fig.~\ref{fig:kappa}, we see that quite small couplings are needed to realize a sufficiently long proton lifetime in both scenarios. 
As $\lambda'$ and $\kappa_1$ are suppressed, all proton decay channels are suppressed due to an increasing SU(5) breaking vev. 
In previous work to avoid small values of $\lambda'$, the GUT scale phases were scanned over to utilize the fact that the Wino and Higgsino-mediated contributions to proton decay would cancel each other. However, for Benchmark point 1, the Higgsino is the lightest supersymmetric particle (LSP) and hence the dark matter candidate. It is also much lighter than the other supersymmetry particles. This strongly suppresses the Higgsino contribution preventing it from canceling with the Wino contribution. As we will see below, the higher-dimensional operators that contribute to the Yukawa couplings can salvage this point even for $\lambda'\sim 1$.

Although the left and right panels of Fig.~\ref{fig:kappa} show some quantitative differences in the behavior of the proton lifetime, the qualitative behavior is quite similar. This is because the sole source of the difference lies in the resulting SU(5)-breaking vev $V$, since these couplings have no significant effect on the Yukawa couplings or soft masses. However, for the scenario with $\kappa_1$, the proton lifetime is long enough to satisfy current experimental limits for a comparatively larger value of $\kappa_1$ than $\lambda'$. This is, however, simply due to the fact that the SU(5)-breaking vev is driven by a higher-dimensional operator. In a naive sense, the higher-dimensional operator can be thought of as an effective tree-level coupling suppressed by $\kappa_1 V/M_P$. This intuition is sufficient to explain the shift in the vev for a given $\kappa_1$ as well as the rough shift in the proton lifetime seen in the figures.

\subsection{The Yukawa sector in the CMSSM}
We now examine the effects of the Planck-suppressed operators that contribute to the Yukawa couplings of the colored Higgs. We will ignore all other higher-dimensional operators except that proportional to $c$, since it allows $\lambda$ and $\lambda'$ to be chosen freely.

In the absence of any Planck-suppressed operators, the colored Higgs boson Yukawa couplings are determined by the Standard Model Yukawa couplings $f_{u,i}$, and $f_{d,i}$ or $f_{e,i}$, as can be seen in Eq.~\eqref{eq:fqqfue} and Eq.~\eqref{eq:fqlfud}. However, allowing the $c_{\Delta h,i}$ to be non-zero breaks this correspondence. In this Section, we examine explicitly the effect on the proton lifetime of these additional contributions to the colored Higgs Yukawa couplings.  

Although there are four higher-dimensional operators contributing to the colored Higgs Yukawa couplings, only three of them can be used to alter the proton decay signatures.  This is due to the fact that $c_{\Delta h,2}$ is used to explain the difference in the lepton and down type quark Yukawa couplings.  This means that only three of the Yukawa couplings can be suppressed. Examining Eq.~\eqref{eq:fqlfud}, it is clear that $f_{QL}$ or $f_{UD}$ cannot both be suppressed. Because the Wino-mediated contribution to the proton decay width, which is proportional to $C_{5L}$ as seen in Eq.~\eqref{eq:CHCW}, is usually the dominant contribution, we choose to suppress $f_{QL}$ and leave $f_{UD}$ alone. To facilitate this suppression strategy, we reparameterize the $c_{\Delta h,i}$ as follows 
\begin{align}
    (r_{ql})_i &= 1+\frac{5}{\sqrt{2}}\frac{(c_{\Delta h,1})_i}{f_{e,i}}R~, \nonumber\\
    (r_{ue})_i &= 1+\biggl(5\frac{(c_{\Delta h,4})_i}{f_{u,i}}-\frac{5}{4}\frac{(c_{\Delta h,3})_i}{f_{u,i}}\biggr)R~, \nonumber\\
    (r_{qq})_i &= 1+\biggl(7\frac{(c_{\Delta h,4})_i}{f_{u,i}}-\frac{(c_{\Delta h,3})_i}{f_{u,i}}\biggr)R~, 
\end{align}
where we use the same basis as in Eq. (\ref{eq:cdh1}-\ref{eq:cdh4}) for the $(c_{\Delta h, k})_{ij}$. We point out at this point that the $r$'s in the above equations are in a naive sense a measure of the needed tuning to suppress the Yukawa coupling.   Again, $c_{\Delta h, 2}$ does not appear here, since it is used to account for the splitting in the lepton and down type quark Yukawa couplings in the Standard Model. Using these definitions, the colored Higgs Yukawa couplings can then be defined as follows  
\begin{align}
    f_{QQ,i}&=(r_{qq})_if_{u,i}~,\nonumber\\
    f_{UE,i}&=(r_{ue})_if_{u,i}~,\nonumber\\
    f_{QL,i}&=(r_{ql})_if_{e,i}~,\nonumber\\
    f_{UD,i}&=f_{d,i}-f_{e,i} \bigl( 1-(r_{ql})_i \bigr) ~.
\end{align}

We now examine the effects of these additional degrees of freedom on the parameter space of minimal SU(5). 
In particular, we analyze the effect on the proton lifetime when the colored Higgs Yukawa couplings are varied. 
We first look at the variation in the proton lifetime for a given supersymmetric spectrum.  
In addition to Benchmark point 1, discussed above, we also consider a second benchmark point
\begin{itemize}
    \item Point 2: $m_0=14.15$~TeV, $m_{1/2}=9.8$~TeV, $A_0=3m_0$, $\tan \beta=5$, $\mu>0$. $\lambda=\lambda'=1$~.
\end{itemize}
For this point, the LSP is a Bino and its relic density is controlled by stop co-annihilations. 
Table~\ref{tab:point2}
summarizes the inputs, derived GUT-scale quantities, MSSM parameters and observables for this benchmark point.

\begin{table}[ht]
  \centering
  \begin{tabular}{cccc}
  \hline \hline
  \multicolumn{4}{c}{Inputs}\\
  \hline
  $m_{1/2}=9.8$~TeV & $m_0=14.15$~TeV & $A_0/m_0=3$ & $\tan{\beta}=5$ \\
  $\lambda=1$ & $\lambda'=1$ & & \\
  \hline
  \multicolumn{4}{c}{GUT-scale parameters (masses in units of $10^{16}~$GeV)}\\
  \hline
  $M_{\rm GUT}=0.685$& $M_{H_C}=1.257$ & $M_\Sigma=0.628$ & $M_X=0.862$ \\
  $V=0.251$ & $g_5=0.686$ & $c=-0.481$ & \\
  \hline
  \multicolumn{4}{c}{MSSM parameters (masses in units of TeV)}\\
  \hline
  $m_\chi=4.78$ & $m_{\tilde{t}_1}=4.84$ & $m_{\tilde{g}}=19.15$ & $m_{\chi_2}=4.84$ \\
  $m_{A}=24.21$  
  & $\mu=18.87$ & $m_{\tilde{\ell}_L}=15.35$& $m_{\tilde{\ell}_R}=14.56$ \\
  $m_{\tilde{\tau}_1}=14.44$ & $m_{\tilde{q}_L}=20.9$ & $m_{\tilde{d}_R}=20.3$ & $m_{\tilde{t}_2}=15.78$ \\
  $A_t=32.5$ & $A_b=65.6$ & $B=-11.43$ & \\
  \hline
  \multicolumn{4}{c}{Observables}\\
  \hline
  $\Omega_\chi h^2=0.099$ & $m_h=124.3$~GeV & & \\
  \hline \hline
  \end{tabular}
  \caption{Benchmark point 2 parameters.  The observables in the final line are the dark matter relic density and the Higgs mass calculated with {\tt FeynHiggs 2.18.1.} }
  \label{tab:point2}
\end{table}

\begin{figure}[ht!]
    \centering
    \includegraphics[width=0.40\linewidth]{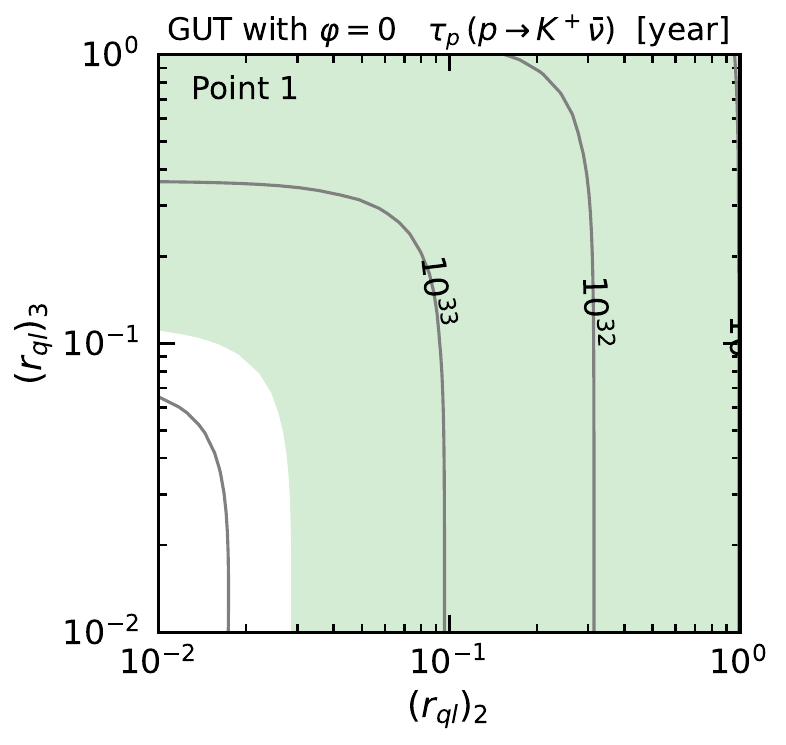}
    \includegraphics[width=0.40\linewidth]{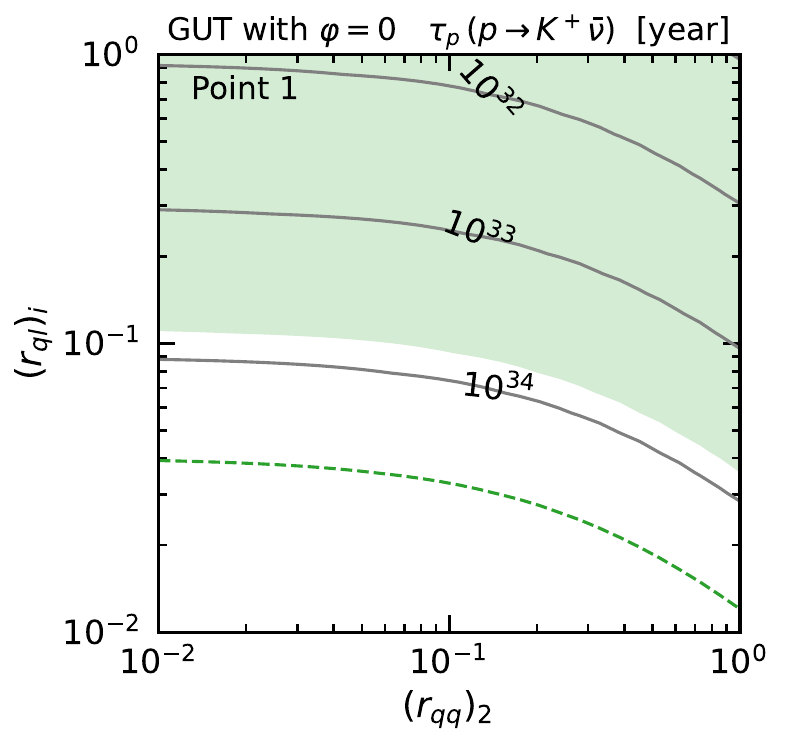}
    \includegraphics[width=0.40\linewidth]{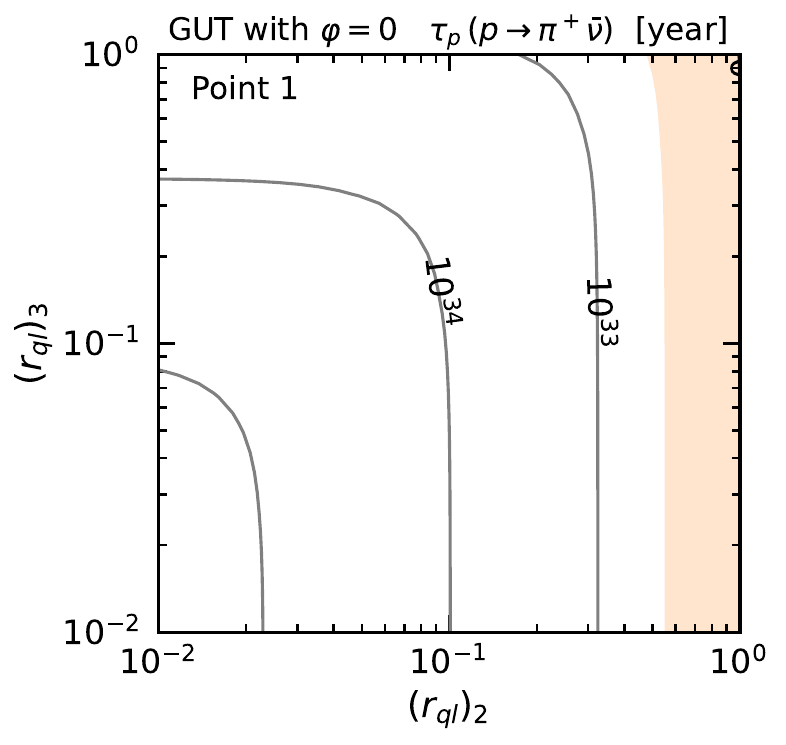}
    \includegraphics[width=0.40\linewidth]{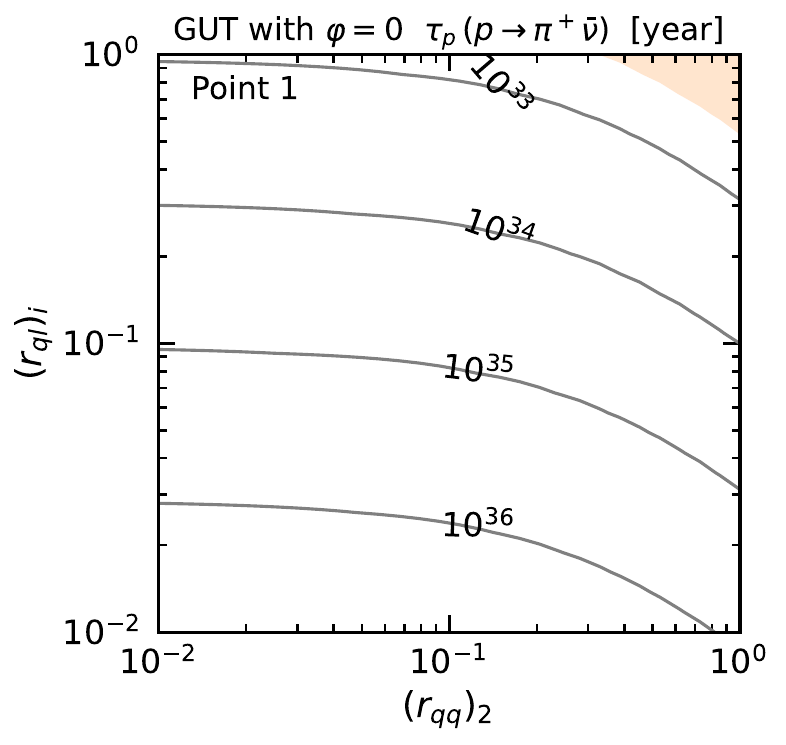}
    \includegraphics[width=0.40\linewidth]{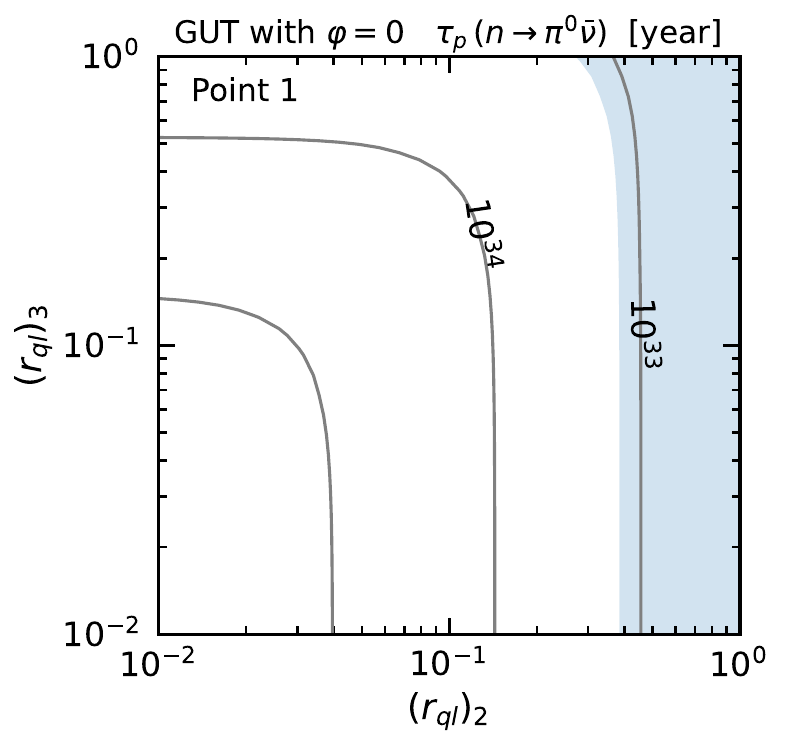}
    \includegraphics[width=0.40\linewidth]{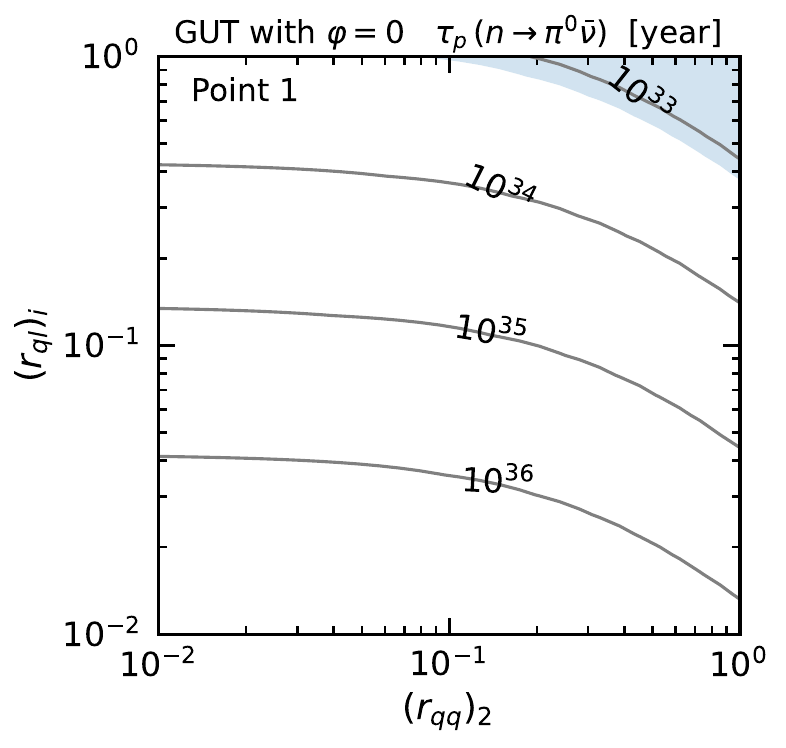}
    \vspace{-5mm}
    \caption{Left panels: dependences of proton lifetimes on $((r_{ql})_2, (r_{ql})_3)$ for point 1 with $(r_{ql})_1=(r_{qq})_i=(r_{ue})_i=1$. Right panels: dependences of proton lifetimes  on $(r_{qq})_{2}, (r_{ql})_{i=1,2,3})$ for point 1 with $(r_{qq})_{i=1,3},(r_{ue})_{i=1,2,3}=1$. In both cases it is assumed that the GUT phases $\varphi_i=0$. The shaded areas show the current constraints from $p \to K^+ \bar \nu$ (top panels), $p \to \pi^+ \bar \nu$ (middle panels) and $n \to \pi^0 \bar \nu$ (bottom panels).}
    \label{fig:fql}
\end{figure}

Figure~\ref{fig:fql} explores the dependence of selected nucleon decay lifetimes (solid lines) for Benchmark point 1, with $\lambda'=\lambda=1$ and $\kappa_1=0$, in the $((r_{ql})_2, (r_{ql})_3)$ plane (left plots) assuming $(r_{ql})_1= (r_{qq})_{i=1,2,3}=(r_{ue})_{i=1,2,3}=1$, and in the
$((r_{qq})_{2}, (r_{ql})_{i=1,2,3})$ plane (right plots) assuming $(r_{qq})_{i=1,3} = (r_{ue})_{i=1,2,3}=1$.  In both cases, we assume that the GUT phases $\varphi_i = 0$. 
From top to bottom, we show the lifetimes $\tau(p\to K^+\bar{\nu})$, $\tau(p\to \pi^+\bar{\nu})$, and $\tau(n\to \pi^0\bar{\nu})$, comparing them to the
experimental limits. The shaded areas are the current decay limits for each respective channel, and the dashed line in the upper right panel shows the $p\to K^+\bar{\nu}$ prospective sensitivity of Hyper-Kamiokande. In the upper left panel, the Hyper-Kamiokande sensitivity is strong enough to explore all of the displayed parameter space. 
We see that in both cases $p \to K^+ \bar \nu$  searches provide the strongest constraints, with the bound on $(r_{ql})_2$ being stronger than that on $(r_{ql})_3$ and the bound on $(r_{ql})_i$ being stronger than that on $(r_{qq})_2$.  As can be seen in the left plots of this figure, $(r_{ql})_3 \lesssim 0.1$ requires less fine tuning than $(r_{ql})_2$, which is constrained to $\lesssim 0.03$.~\footnote{We do not show results for $(r_{ql})_1$ in this figure, as it is much less efficient at enhancing the proton lifetime, due to the smallness of electron Yukawa coupling $f_{e,1}$.}.

In the left panels of Fig.~\ref{fig:fql}, we examine the proton lifetime dependence on $f_{QL}$ using $r_{ql}$ as a surrogate to highlight the needed fine tuning, if this was the only coupling to be suppressed. This choice was made based on the fact that the Wino-mediated contribution to the proton decay rate is usually the largest, and can be suppressed by this coupling. As is clear from the left panels of Fig.~\ref{fig:fql}, this strategy works but does require some degree of tuning of both $f_{QL,2}$ and $f_{QL,3}$.
The amount of tuning can be reduced by considering the suppression of other colored Higgs Yukawa couplings that contribute to proton decay. Since the suppression of $f_{QL}$ has used all the freedom in the $c_{\Delta h,1}$ couplings, we are only allowed to suppress $f_{QQ,i}$ and $f_{UE,i}$ using $r_{QQ}$ and $r_{UE}$.  However, since the third-generation components of these colored Higgs boson Yukawa couplings receive contributions proportional to the top Yukawa coupling, $(f_u)_3$, the suppression of $(f_{QQ})_{3}$ and $(f_{UE})_{3}$ requires values of $(c_{\Delta h,4}^S)_{3}$ that are rather large. (We discuss the sizes of the $c_{\Delta h, i}$ in more details in the next Section.) To avoid these large couplings~\footnote{As we discuss in the following Section, these large couplings may not be problematic. However, to show that large couplings can be avoided we do not consider $f_{QQ,3}$ here.}, we have only considered the suppression of $f_{QQ,2}$ in the right panels of Fig.~\ref{fig:fql}~\footnote{ Of the couplings in Eq.~\eqref{eq:CHCW}, the leading order contributions to proton decay come from $(f_{UE})_3$ and $(f_{QQ})_{i=2,3}$.}.

In the $((r_{qq})_{2}, (r_{ql})_{i=1,2,3})$ planes shown in the right panels of Fig.~\ref{fig:fql}, we see again that the
$p \to K^+ {\bar \nu}$ decay channel provides the strongest constraint on the parameter space. For low values of $(r_{qq})_2$ we see that $(r_{ql})_i \lesssim 0.1$ and the limit improves to $(r_{ql})_i \lesssim 0.03$ for $(r_{qq})_2 \simeq 1$. Hyper-Kamiokande will improve these sensitivities by roughly a factor of 3 as seen by the 
dashed curve in the upper right panel.

So far in this Section, we have only considered Benchmark point 1.  As remarked earlier, this point has a Higgsino-like LSP which is a viable dark matter candidate. For this point the $\mu$ parameter is quite suppressed compared to the other supersymmetry parameters.  This feature already suppresses the dimension-5 Higgsino-dependent proton decay amplitudes, so we need only consider the possible suppression of the Wino-mediated contributions to the dimension-5 proton decay width. This additional suppression is indeed required for a viable proton lifetime to be found in the simplified scenarios we consider.~\footnote{We consider later the effects of the phases of these couplings, which allow stop coannihilation-type dark matter candidates to be viable even for $\lambda'\sim 1$.}   

We now examine Benchmark point 2, which has a Bino Dark matter candidate and a Higgsino mass, $\mu$, which is much larger than the Wino mass. This means that the Higgsino-mediated contribution to proton decay cannot be suppressed by tuning it against the Wino-mediated contribution. Because of this, the proton lifetime tends to be too short if $\lambda=\lambda'=1$\footnote{If we take $\lambda'\ll 1$, the SU(5) breaking vev becomes much larger and the proton lifetime can be enhanced.}.  This is why the previous figures focused on Benchmark point 1.

We now show that Benchmark point 2 may still be viable even with the assumption that only $r_{ql}$ is suppressed, if we vary $\lambda$ and $\lambda'$. 
We display in Fig.~\ref{fig:f_lamlamp} the lower limits on $\lambda$ (left panels) and upper limits on $\lambda'$ (right panels) as functions of $(r_{ql})_i = (r_{ql})_1=(r_{ql})_2=(r_{ql})_3$ assuming $\lambda'=1$ (left panels) and  $\lambda=1$ (right panels). We show their dependences for both Benchmark point 1 (upper plots) and Benchmark point 2 (lower plots). The solid lines correspond to the current experimental bounds and the dashed lines correspond to projected proton decay sensitivities of Hyper-Kamiokande. The bounds on $\lambda'$ and $\lambda$ are very different due to the fact that $M_{H_C}$ scales proportional to $\lambda$ and inversely proportional to some power of $\lambda'$. These figures also show that $r_{ql}$ suppression alone is sufficient for point 1 with $\lambda=\lambda' =1$. However, for point 2 additional suppression beyond $r_{ql}$ is needed to get a viable proton lifetime, i.e., the larger SU(5) breaking vev found for small $\lambda'$ (see Fig.~\ref{fig:kappa}). In the left plots of Fig.~{\ref{fig:f_lamlamp}, we see that $p \to K^+ \bar \nu$ always provides the strongest constraint on $\lambda$. This is mostly the case for the figures in the right panels, although $p \to \pi^0 e^+$ becomes dominant for current constraints and will remain competitive for future constraints for point 1 when $r_{ql} = {\cal O}(10^{-2})$.

\begin{figure}
    \centering
    \includegraphics[width=0.42\linewidth]{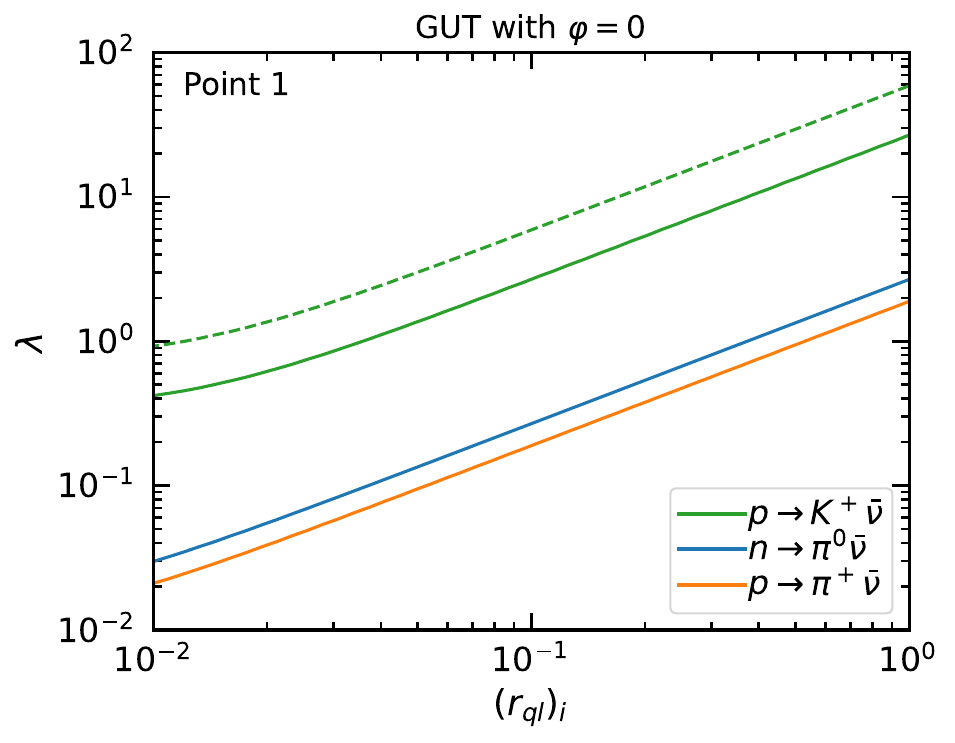}
    \includegraphics[width=0.42\linewidth]{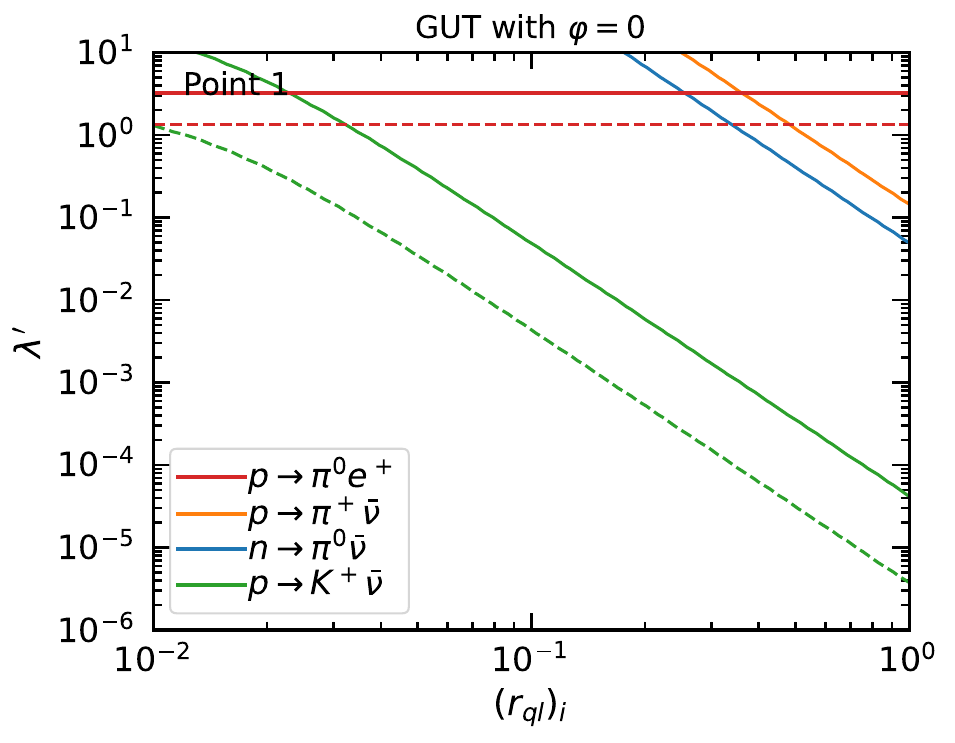}
    \includegraphics[width=0.42\linewidth]{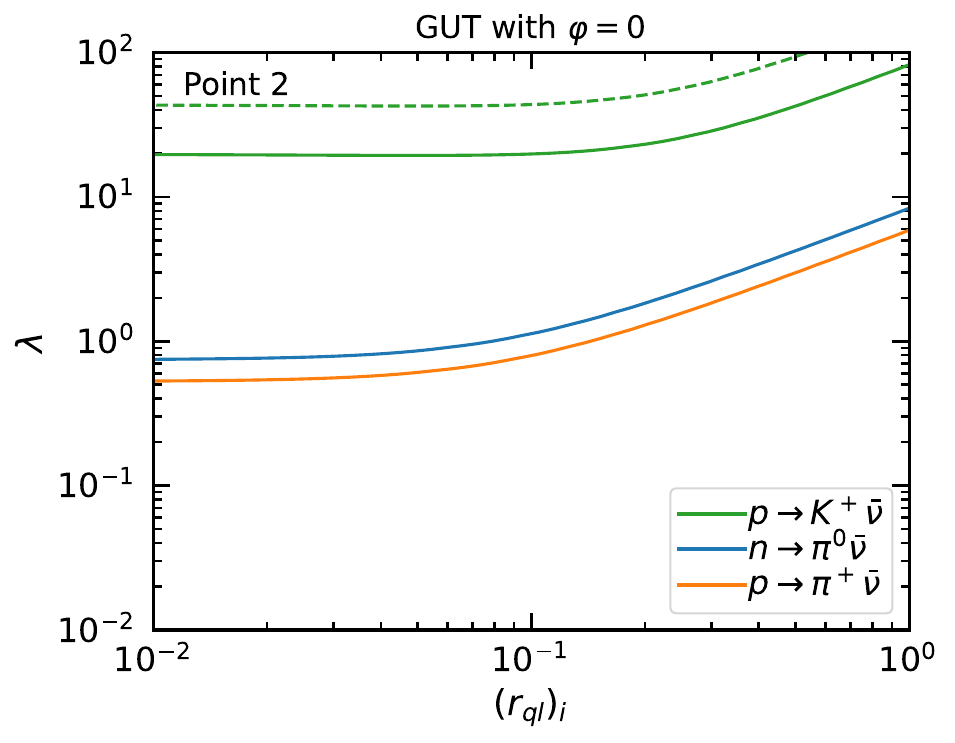}
    \includegraphics[width=0.42\linewidth]{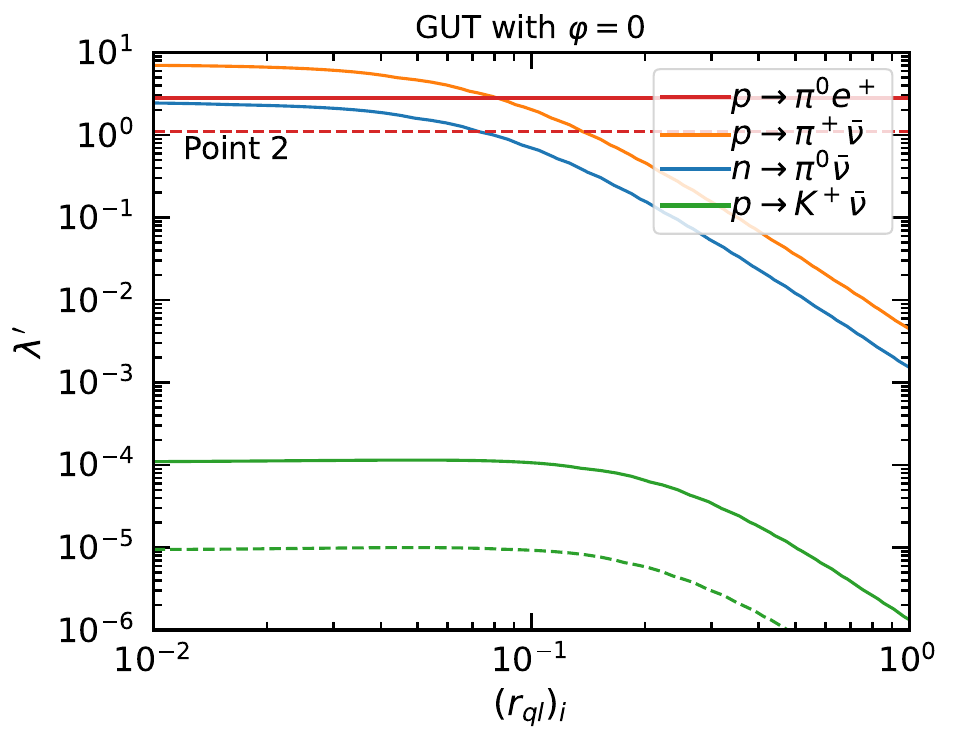}
    \caption{Lower limits on $\lambda$ as functions of $(r_{ql})_i$ assuming $\lambda'=1$ (left panels) and upper limits on $\lambda'$ assuming $\lambda=1$ as functions of $(r_{ql})_i$ (right panels), for Benchmark point 1 (upper plots) and Benchmark point 2 (lower plots). We assume in all cases that the GUT phases $\varphi_i = 0$ and $(r_{qq})_i = (r_{ue})_i = 1$~.
    The solid lines correspond to the current experimental bounds and the dashed lines correspond to future proton decay sensitivities of Hyper-Kamiokande. 
    }
    \label{fig:f_lamlamp}
\end{figure}

\subsection{Dependence on the Planck-suppressed Yukawa couplings $c_{\Delta h,i}$}
\label{depend}

In this Section, we determine the order of magnitude of the couplings of the Planck-suppressed operators needed to significantly reduce the proton decay width in the CMSSM scenarios for our benchmark points. Using the parameter set for Benchmark point 1,
we determine the sizes of the coefficients of the the Planck-suppressed operators needed to cancel the colored Higgs Yukawa couplings. The approximate value 
for these couplings is as follows~\footnote{Note that the values of $c_{\Delta h,i}$ estimated here scale with $v_0$. For all numerical calculations, we calculate $v_0$ using the matching conditions and it tends to be ${\mathcal O}(10^{16}~{\rm GeV})$. Hence, the values of $c_{\Delta h,i}$ presented here are merely representative and the true value will deviate slightly depending on the exact point we consider.},
\begin{align}
    c_{\Delta h,2}=\frac{\sqrt{2}}{5}\frac{1}{R}(f_d-f_e) = 
    \begin{pmatrix}
        0.00072~\bigl(\frac{(f_d)_1-(f_e)_1}{1.06\times 10^{-5}}\bigr) \\
        -0.095~\bigl(\frac{(f_d)_2-(f_e)_2}{-1.4\times 10^{-3}}\bigr) \\
        -0.54~\bigl(\frac{(f_d)_3-(f_e)_3}{-8\times 10^{-3}}\bigr)
    \end{pmatrix}
    \biggl( \frac{10^{16} ~{\rm GeV}}{v_0}\biggr)~,
    \label{eq:ch2}
\end{align}
\begin{align}
    c_{\Delta h,1} &=-\frac{\sqrt{2}}{5}\frac{1}{R}f_e =
    \begin{pmatrix}
        -0.00057~\bigl(\frac{(f_e)_1}{0.84\times 10^{-5}}\bigr) \\
        -0.12~\bigl(\frac{(f_e)_2}{1.74\times 10^{-3}}\bigr) \\
        -1.97~\bigl(\frac{(f_e)_3}{0.029}\bigr)
    \end{pmatrix}
    \biggl( \frac{10^{16} ~{\rm GeV}}{v_0}\biggr) ~,\\
    c_{\Delta h,3} &=\frac{8}{15}\frac{1}{R}f_u =
    \begin{pmatrix}
        ~{\cal O}(10^{-4}) \\
        0.24~\bigl(\frac{(f_u)_2}{1.845\times 10^{-3}}\bigr) \\
        70~\bigl(\frac{(f_u)_3}{0.546}\bigr)
    \end{pmatrix}
    \biggl( \frac{10^{16} ~{\rm GeV}}{v_0}\biggr) ~,\\
    c_{\Delta h,4} &=-\frac{1}{15}\frac{1}{R}f_u =
    \begin{pmatrix}
        ~-{\cal O}(10^{-5}) \\
        -0.0295~\bigl(\frac{(f_u)_2}{1.845\times 10^{-3}}\bigr) \\
        -8.7~\bigl(\frac{(f_u)_3}{0.546}\bigr)
    \end{pmatrix}
    \biggl( \frac{10^{16} ~{\rm GeV}}{v_0}\biggr) ~.
    \label{eq:cyukawa}
\end{align}
\SH{The parameter $c_{\Delta h,2}$ is used to accommodate the difference between lepton and down quark Yukawa couplings, where we consider two-loop RGEs for gauge couplings and Yukawa couplings and include one-loop threshold corrections. The contribution of supersymmetry threshold corrections to the Yukawa splitting is taken into account in the computation.}
The expressions for $c_{\Delta h, 1}, c_{\Delta h, 3}$, $c_{\Delta h,4}$ are found by solving Eq.~(\ref{eq:fqqfue}) and Eq.~(\ref{eq:fqlfud}) for $f_{Q L,_i}=0$, $f_{UE,_i}=0$, and $f_{QQ,_i}=0$. Although some of these couplings may seem large, they are the coefficients of non-renormalizable operators. The important thing for these operators is that the effective Yukawa coupling they generate is perturbative. Since all these operators come with a $V/M_P\sim 10^{-2}$ suppression, even $(c_{\Delta h, 3})_3\sim 70$ could be a reasonable value. 
Furthermore, the apparent large values of the Wilson coefficients may correspond to the possibility that the scale at which these operators are induced is somewhat below the Planck scale and/or that these operators are generated through the exchange of many fields.
Nevertheless, we focus on a more standard scenario where the coefficients of the higher-dimensional operators are closer to unity.

Another viable path for suppressing proton decay is to let $f_{QQ}$ be unsuppressed, while the suppression of only $f_{QL}$ and $f_{UE}$ is used to reduce the Wino- and Higgsino-mediated contributions to proton decay, respectively. This can be done by tuning $c_{\Delta h, 1}$ to suppress $f_{QL}$ and $c_{\Delta h, 4}$ to suppress $f_{UE}$.
This allows us to ignore $c_{\Delta h,3}$, which had a coefficient of $70$, and focus on $c_{\Delta h,4}$.  In this case we need
\begin{align}
    c_{\Delta h,4} &=-\frac{1}{5}\frac{1}{R}f_u  =
    \begin{pmatrix}
        ~-{\cal O}(10^{-5}) \\
        -0.088~\bigl(\frac{(f_u)_2}{1.845\times 10^{-3}}\bigr) \\
        -26~\bigl(\frac{(f_u)_3}{0.546}\bigr)
    \end{pmatrix}
    \biggl( \frac{10^{16} ~{\rm GeV}}{v_0}\biggr)~, \nonumber\\
    c_{\Delta h,3} &\ll \frac{1}{R}f_u~.
\end{align}
Although this still requires a coefficient to  be nominally larger than one, it nevertheless 
corresponds to a perturbative Yukawa coupling in the effective theory. 

In order to examine the range of the couplings of the Planck-suppressed operators that can give a viable proton lifetime, we again consider Benchmark point 1 and plot in Fig.~\ref{fig:c1i_new} planes involving different combinations of the $(c_{\Delta h,1})_i$. In each plane, the missing $(c_{\Delta h,1})_i$ is set to the value that exactly cancels the corresponding $(f_{QL})_i$. In the left panels we have set the GUT phases $\varphi=0$ while in the right panels we adopt the GUT phases that produce the maximum proton lifetime, found from scanning over the phases. In all of the panels, we have set $c_{\Delta h, 2}= c_{\Delta h,3} =c_{\Delta h,4} =0$. The shaded regions show the regions of the planes which are excluded by current experimental limit on $p \to K^+ {\bar \nu}$, and the dashed curves correspond to the expected future limit from Hyper-Kamiokande.  The solid curves correspond to our computed proton lifetime. As can be seen in these plots, the values of $(c_{\Delta h,1})_i$ needed to obtain a viable proton lifetime are not very large and can be varied by an amount of ${\cal O}(1)$, even for $\varphi=0$. If the phases are varied, this sensitivity to the magnitudes of the $(c_{\Delta h,1})_i$ is even further reduced.  Furthermore, for a judicious choice of couplings, it is possible to make the proton lifetime (for this decay mode) arbitrarily long. 

\begin{figure}[ht!]
    \centering
    \includegraphics[width=0.34\linewidth]{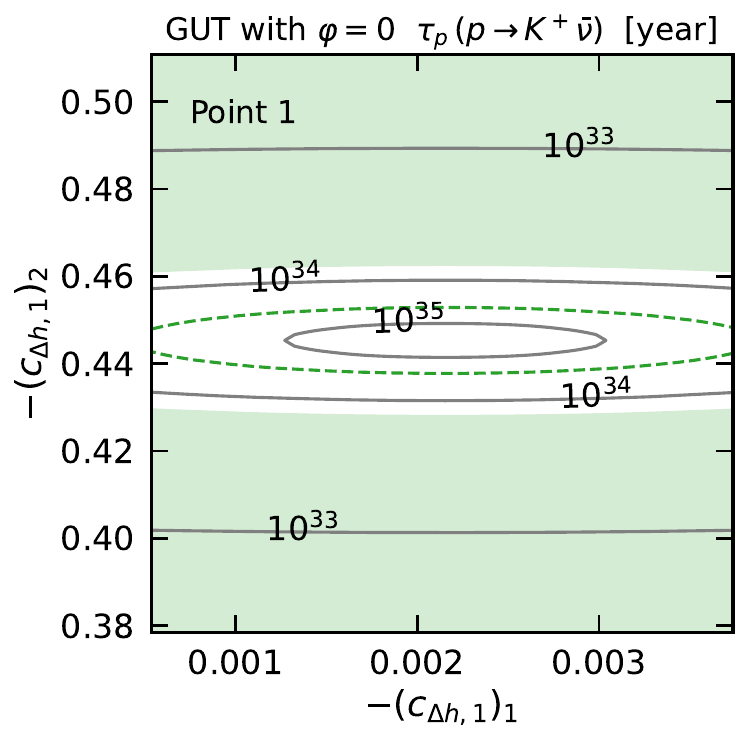}
    \includegraphics[width=0.36\linewidth]{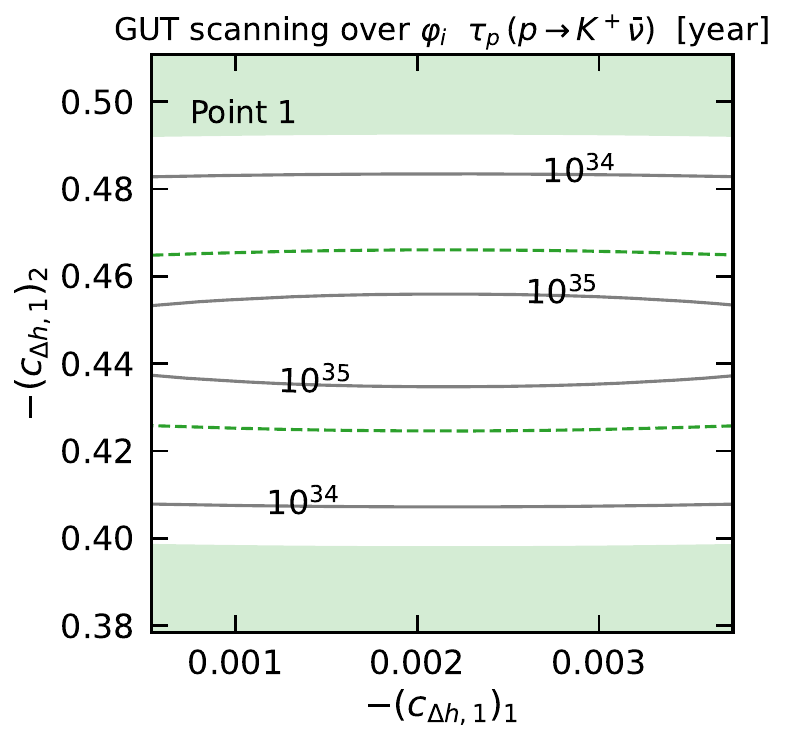}
    \includegraphics[width=0.34\linewidth]{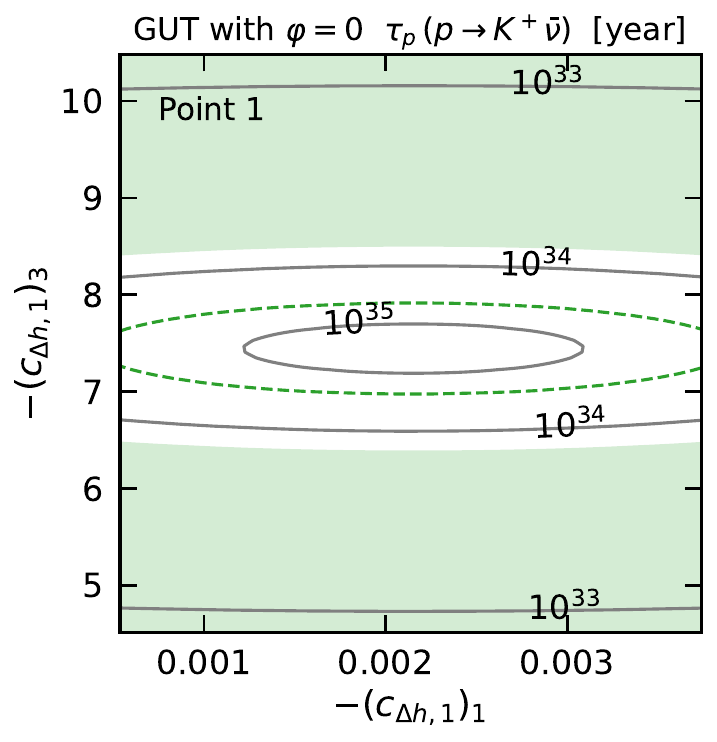}
    \includegraphics[width=0.36\linewidth]{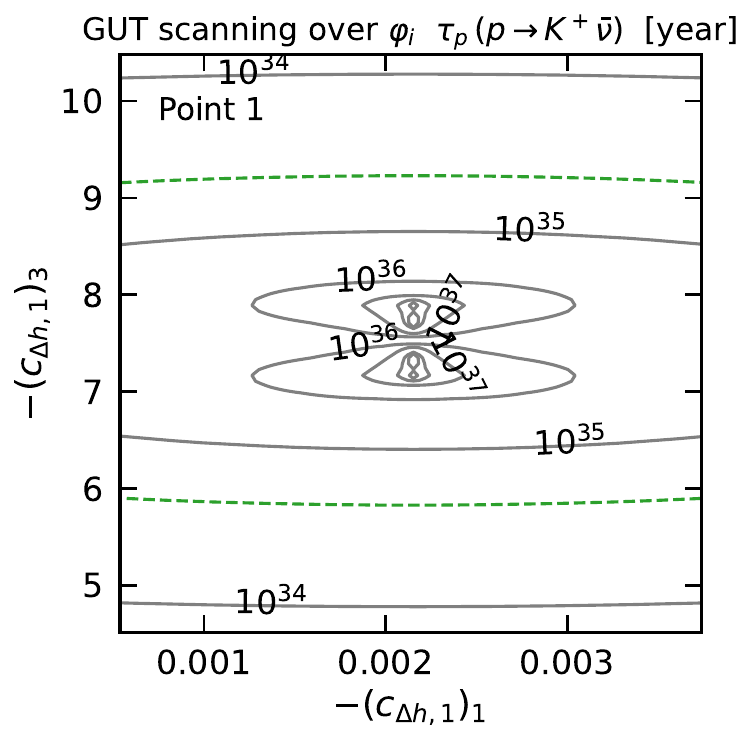}
    \includegraphics[width=0.34\linewidth]{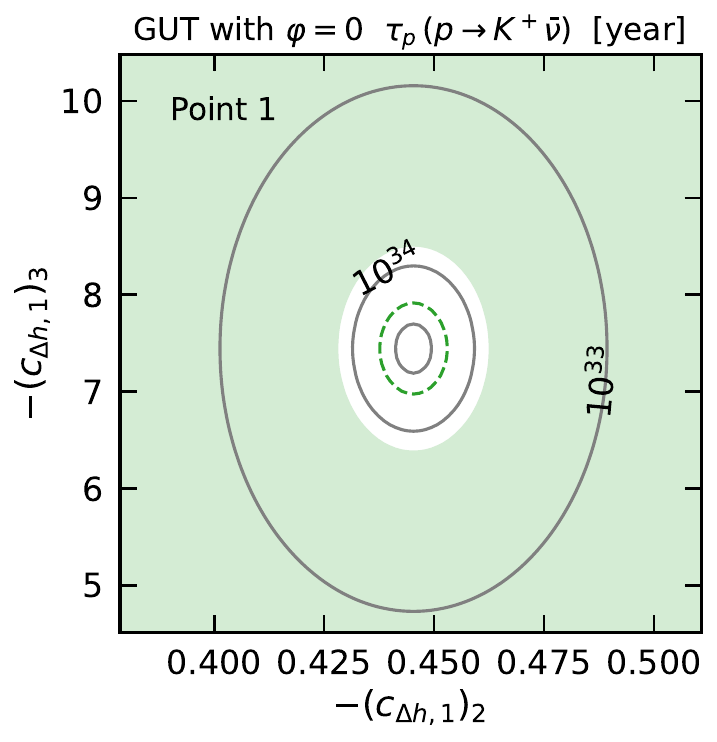}
    \includegraphics[width=0.36\linewidth]{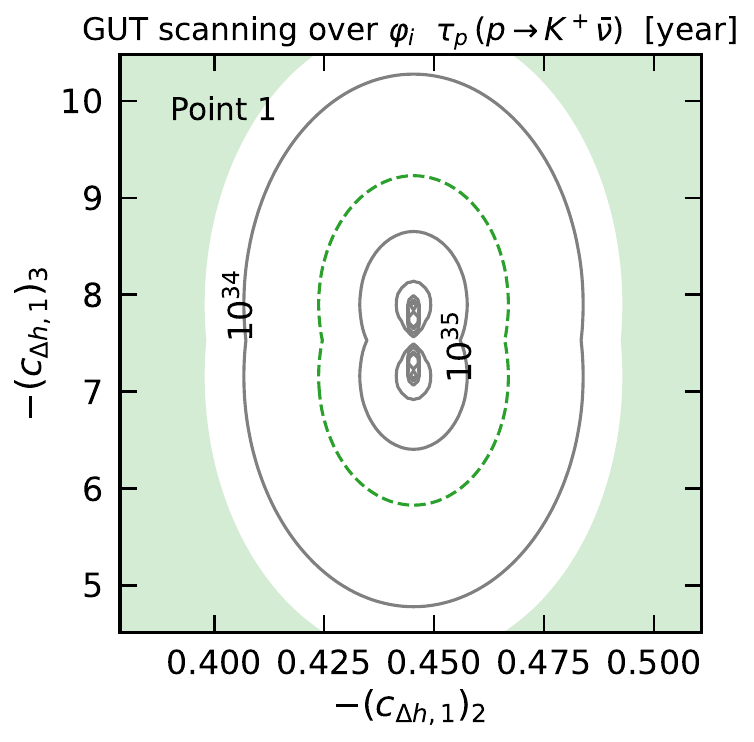}
    \caption{Contours of the $p \to K^+ {\bar \nu}$ proton lifetime for Benchmark point 1 as functions of the $(c_{\Delta h,1})_i$ around the values yielding exact cancellation of $f_{QL,i}$.
    The left panels show the proton lifetimes when the GUT phases $\varphi_i=0$, whereas the right panels show the longest lifetimes found in scans over the phases $\varphi_i$. In the $((c_{\Delta h,1})_1, (c_{\Delta h,1})_2)$ plane, $(c_{\Delta h,1})_3$ is fixed to the value that yields an exact cancellation of $f_{QL,3}$, i.e., $(c_{\Delta h,1})_3=-\frac{\sqrt{2}}{5}\frac{1}{R}(f_e)_3$, and similarly for the other planes. In addition, we set $c_{\Delta h, 2}= c_{\Delta h,3} =c_{\Delta h,4} =0$. 
}
    \label{fig:c1i_new}
\end{figure}

The main contribution to $p\to K^+ {\bar \nu}$ from Higgsino exchange is associated with the third generation, due to the relative magnitudes of the Yukawa couplings in Eq.~\eqref{eq:CHCW}. When $(c_{\Delta h,1})_i$ is used to suppress the Wino-exchange contribution to proton decay, the Higgsino-exchange contribution remains unaffected. However, if the third-generation contribution to the Wino-mediated proton decay amplitude is suppressed so that it is similar in size to the Higgsino-mediated contribution, cancellations between these contributions can occur for a particular phase choice, which can be identified by scanning over $\varphi_i$. Such a cancellation can reduce significantly nucleon decay rates, as seen in Fig.~\ref{fig:lifetime_c13}. In this figure we plot the proton lifetime for several decay modes as a function of $(c_{\Delta h,1})_3$ with $(c_{\Delta h,1})_{1,2}$ chosen such that $f_{QL,i=1,2}=0$. As seen in the right-hand panel of this figure, scanning over the GUT phases manifests the potential significance of the higher-dimensional operators. We see in Fig.~\ref{fig:lifetime_c13} that each of the dimension-5-mediated nucleon decay modes is affected similarly, and that generically the constraint from $p\to K^+\nu$ is the most affected.~\footnote{However, as seen in the figure, this statement breaks down in the case of extreme fine tuning of the $c_{\Delta h,i}$, but our focus is on the minimal amount of fine-tuning that yields an allowed proton lifetime.}

\begin{figure}[ht]
    \centering
    \includegraphics[width=0.45\linewidth]{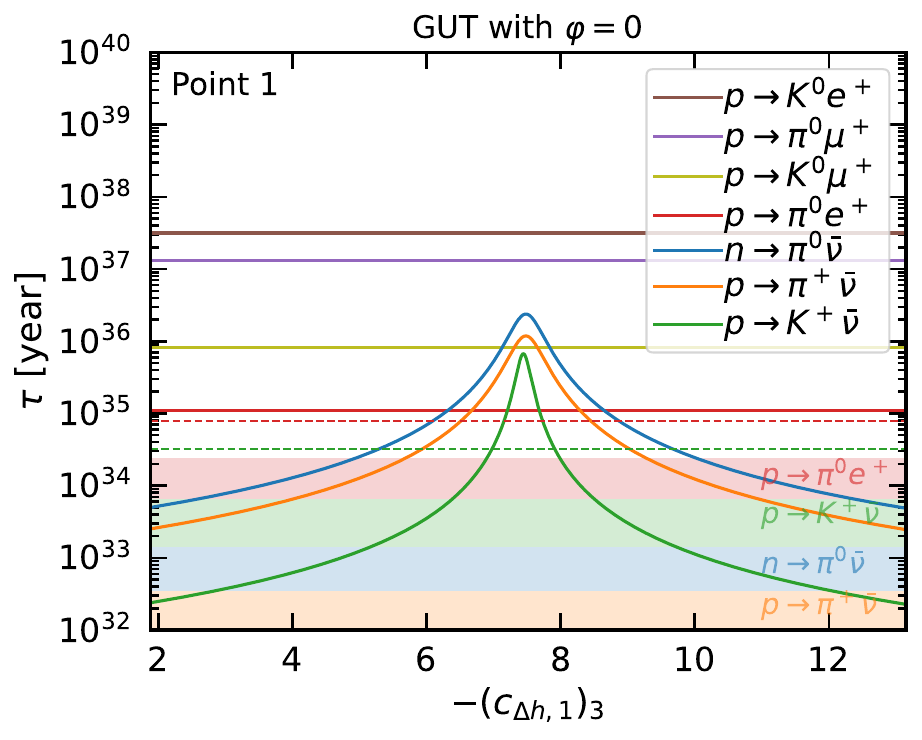}
    \includegraphics[width=0.45\linewidth]{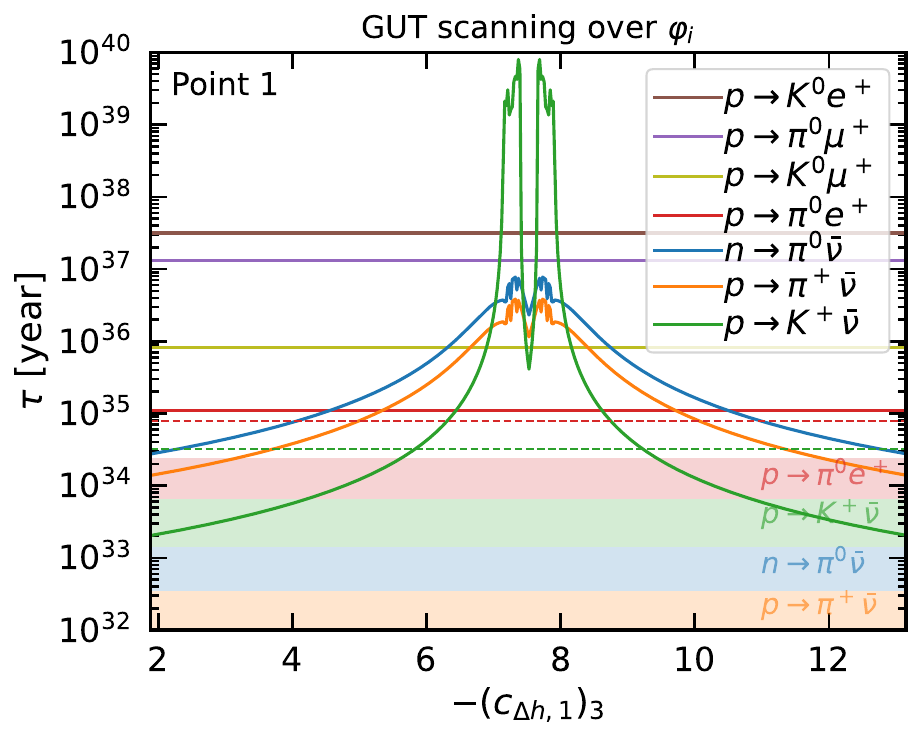}
    \caption{Contour plots of the proton lifetime for Benchmark point 1 as functions of $(c_{\Delta h,1})_3$.
    The values of $(c_{1,\Delta h})_{i=1,2}$  are chosen so that $f_{QL,i=1,2}$ are cancelled. 
    The shaded areas depict the current limit from Super-Kamiokande, while the dashed lines represent the future sensitivities of Hyper-Kamiokande.
    }
    \label{fig:lifetime_c13}
\end{figure}

In all the previous figures, we have taken $\lambda=\lambda'=1$. However, in Fig.~\ref{fig:c12c13} we address the interplay between these couplings and the coefficients $c_{\Delta h,1}$ and, in turn, how this affects proton decay. In the left-hand side of Fig.~\ref{fig:c12c13} we plot the value of $\lambda$ (with $\lambda'=1$) that gives a proton lifetime equal to the experimental bound for various decay modes. Since the proton lifetime scales inversely with some power of $\lambda$, values of $\lambda$ above these lines are not ruled out. Similarly, in the right-hand side of Fig.~\ref{fig:c12c13}, we plot the value of $\lambda'$ (with $\lambda=1$) that gives a proton lifetime equal to the experimental bound for various decay modes. 
We see in the right panels green, blue and orange funnels that contain the viable region of parameter space. The solid lines correspond to the current experimental bounds while the dashed lines correspond to the future proton decay sensitivities of Hyper-Kamiokande.  The upper panels show the sensitivities $\lambda$ and $\lambda'$ with respect to $(c_{\Delta h,1})_2$,
and the lower panels the sensitivities to $(c_{\Delta h,1})_3$.

\begin{figure}[ht]
    \centering
    \includegraphics[width=0.45\linewidth]{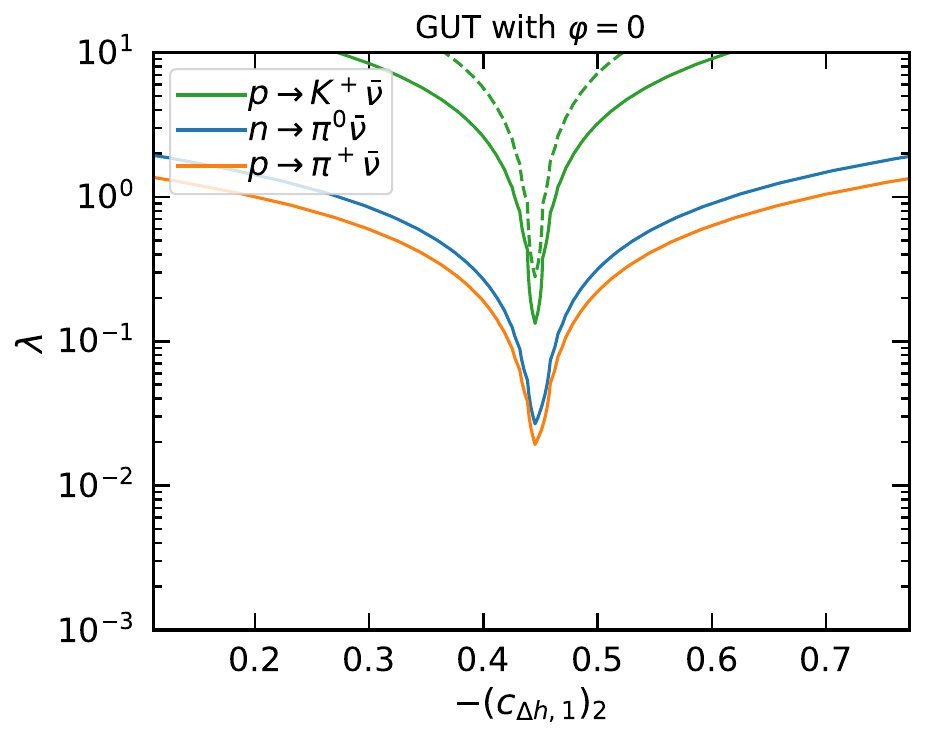}
    \includegraphics[width=0.46\linewidth]{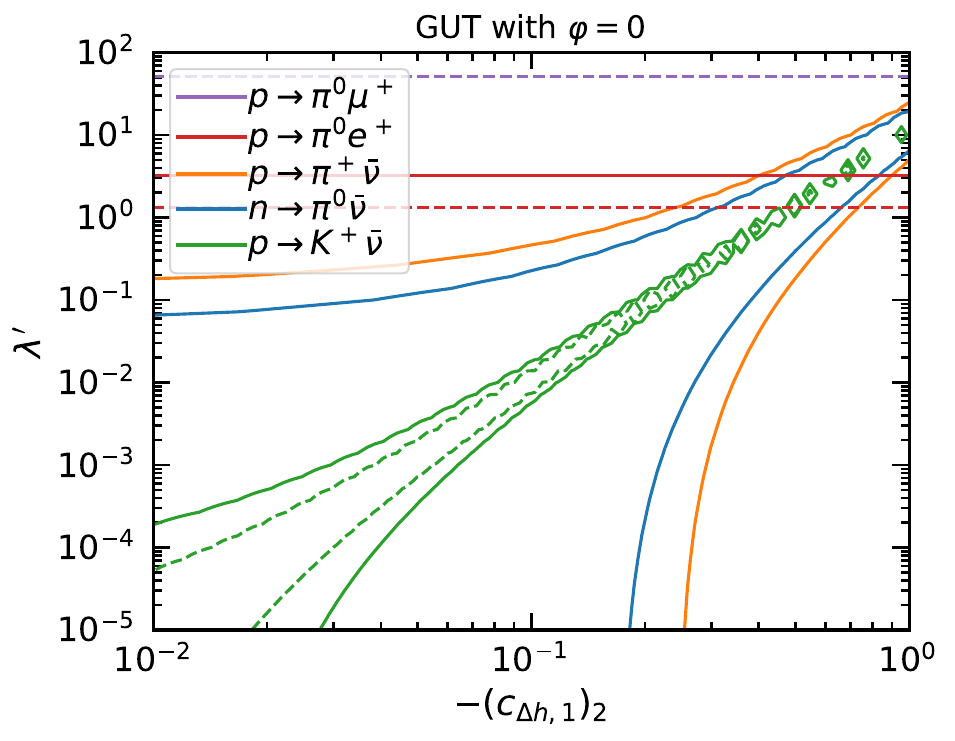}
    \includegraphics[width=0.45\linewidth]{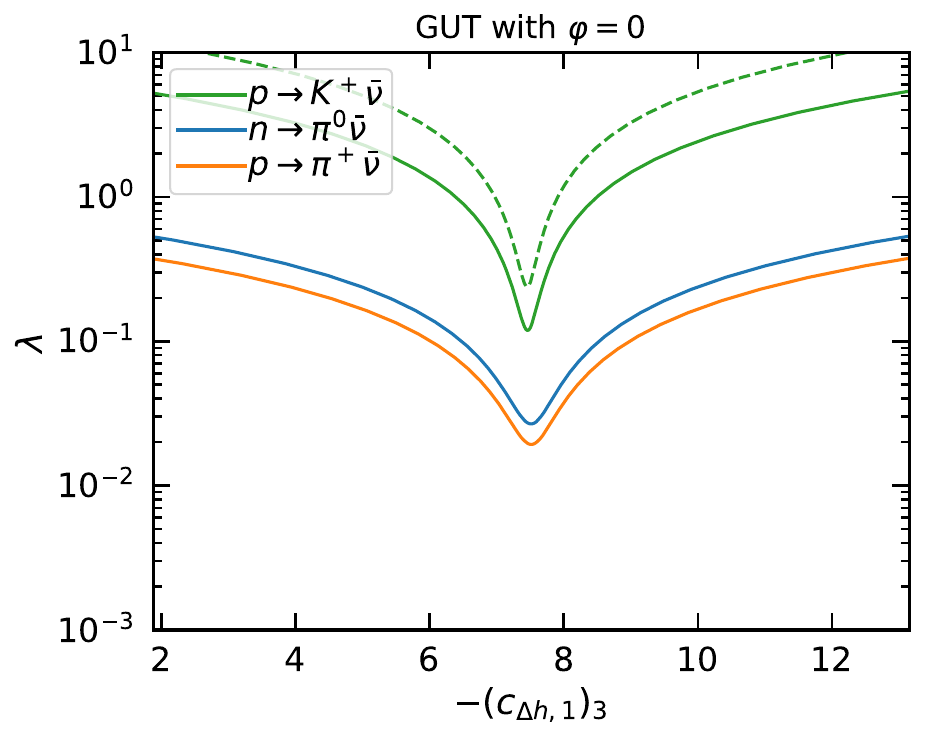}
    \includegraphics[width=0.46\linewidth]{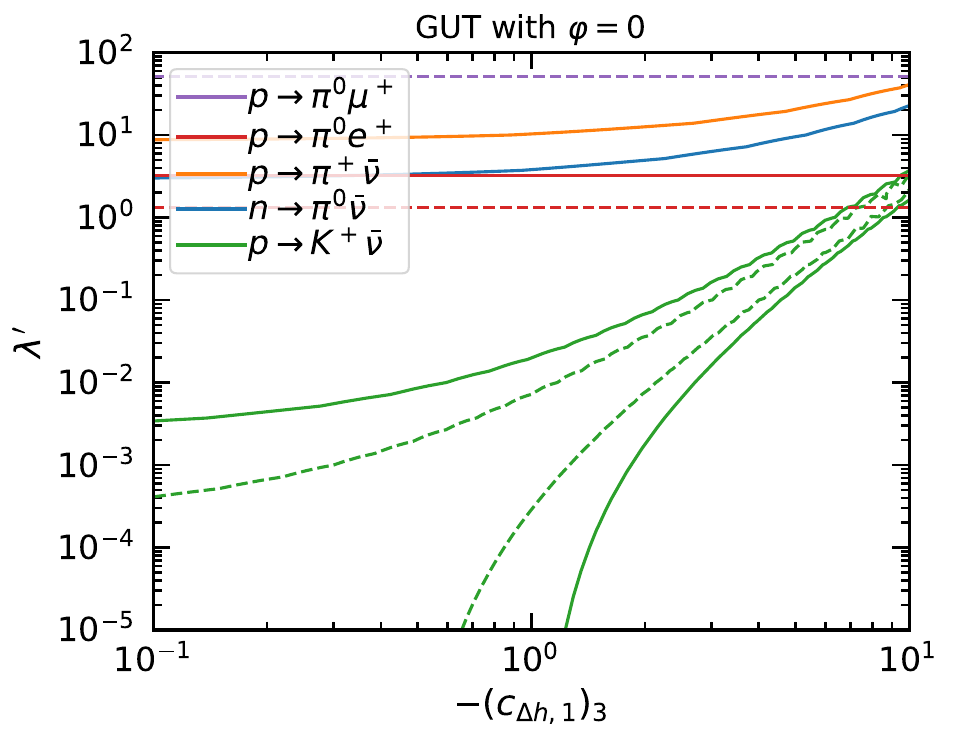}
    \caption{The required value of $\lambda$ (with $\lambda'=1$) (or vice versa) for Benchmark point 1 when the proton lifetimes saturate the experimental bounds. The solid lines correspond to the current experimental bounds while the dashed lines correspond the future proton decay sensitivities of Hyper-Kamiokande. In the $((c_{\Delta h,1})_2, \lambda)$ and $((c_{\Delta h,1})_2, \lambda')$ planes, $(c_{\Delta h,1})_{i=1,3}$  are fixed so that $f_{QL,i=1,3}$ are cancelled. In the $((c_{\Delta h,1})_3, \lambda)$ and $((c_{\Delta h,1})_3, \lambda')$ planes, $(c_{\Delta h,1})_{i=1,2}$  are fixed so that $f_{QL,i=1,2}$ are cancelled.}
    \label{fig:c12c13}
\end{figure}

Finally, we present in Fig.~\ref{fig:m0m12} two examples of ($m_{1/2}, m_0$) planes in the CMSSM. The left panels of Fig.~\ref{fig:m0m12} are for a CMSSM spectrum with $\lambda=\lambda'=1$, $\tan{\beta}=4$, $A_0=0$, $\mu>0$.
That is, these panels correspond to Benchmark point 1, but allowing $m_{1/2}$ and $m_0$ to be free. 
The green star 
shows the position of Benchmark point 1. Similarly, 
the right panels display the scenario where $\lambda=\lambda'=1$, $\tan{\beta}=5$, $A_0=3m_0$, $\mu>0$ as in Benchmark point 2, whose position is also designated by green star. The blue shaded regions have a  dark matter relic density with $0.01 \le \Omega_\chi h^2 \le 2.0$. (We use a large range for  $\Omega_\chi h^2$ to ensure its visibility in the figure.)  The LSP is a higgsino for Benchmark point 1 and a bino for Benchmark point 2.
Electroweak symmetry breaking fails in the purple shaded regions where $\mu^2 <0$ as seen in the upper left corners of the left panels. The lighter stop is the LSP ($m_{t_1} < m_\chi$) in the brown regions seen in the upper left corner of the right panels.  The lighter stau is the LSP in the ochre shaded regions seen in the lower right corners of all panels. In the upper central region of the left panels there are also cyan strips where the LSP is a chargino.

In the left panels, we also set $(c_{\Delta h,2})_i=(c_{\Delta h,3})_i=(c_{\Delta h,4})_i=0$ and use the benchmark point (shown by the green stars) to determine $(c_{\Delta h,1})_i$.
That is, we solve for the values of $(c_{\Delta h,1})_{1,2}$ that cancel $(f_{QL})_{1,2}$. Then we solve for the $(c_{\Delta h,1})_3$, which reduces the Wino-mediated contribution to proton decay to the same size as the Higgsino-mediated contribution. The $(c_{\Delta h,1})_i$ are then fixed to these values across the entire plane. These planes then allow us to see how sensitive this cancellation of the colored Higgs boson Yukawa coupling are to the supersymmetry parameters.
In these panels, we also scan over the GUT scale phases which reduces the supersymmetry parameter dependence to some degree.  However, even if we did not scan over the GUT phases, there would remain some visible regions around the green star with a viable proton lifetime. 

In each panel the region excluded by the current limit on the decay mode considered is shown by color-coded shading. The red dash-dotted lines shows the calculated Higgs mass and the black lines are contours of the lifetime for nucleon decay into the indicated mode. As one can clearly see, the $p\to K^+ {\bar \nu}$ channel constrains most of the visible CMSSM plane, leaving the viable region shaded in blue.  Recall that the area below the blue shading is excluded for a stable LSP as its relic density is too large even though the proton lifetime is sufficiently long. We note that in each panel of Fig.~\ref{fig:m0m12} we have chosen the value of $c_{\Delta h,1}$ at a specific point, either Benchmark point 1 or Benchmark point 2. The figure shows how the proton lifetime varies for fixed $c_{\Delta h,1}$~. The shaded regions would shift for different $c_{\Delta h,1}$~, changing the allowed regions of the $(m_{1/2}, m_0)$ plane.

\begin{figure}
    \centering
    \includegraphics[width=0.35\linewidth]{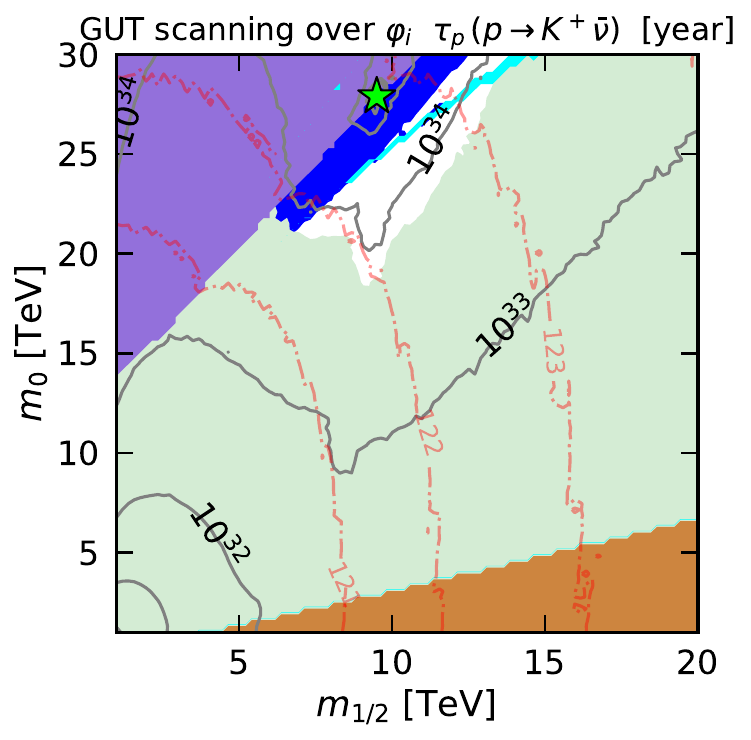}
    \includegraphics[width=0.35\linewidth]{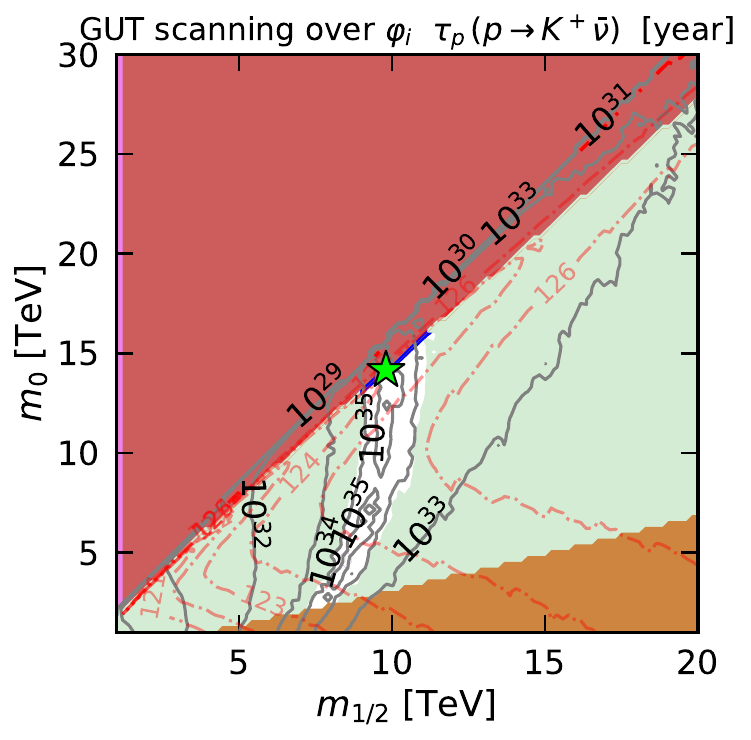}
    \vspace{-2mm}
    \includegraphics[width=0.35\linewidth]{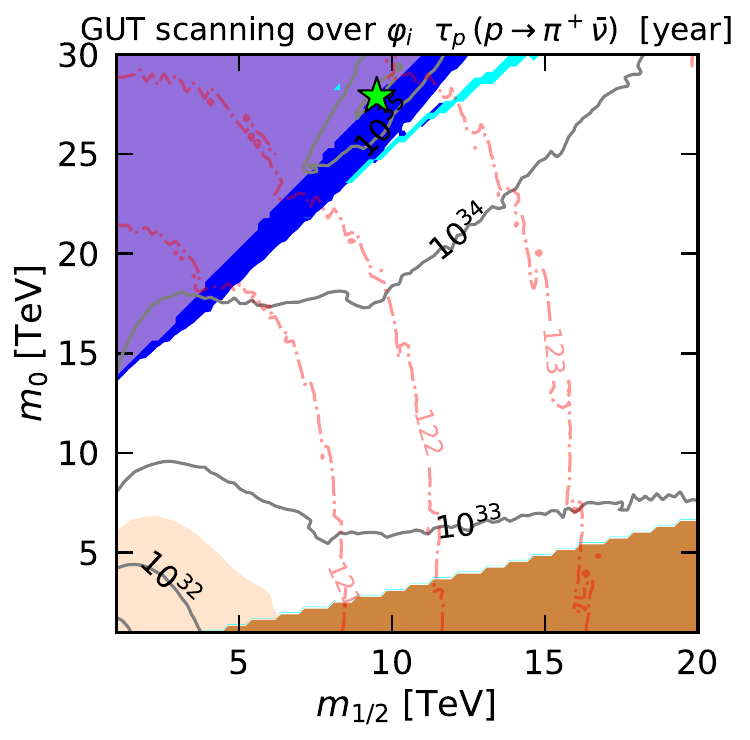}
    \includegraphics[width=0.35\linewidth]{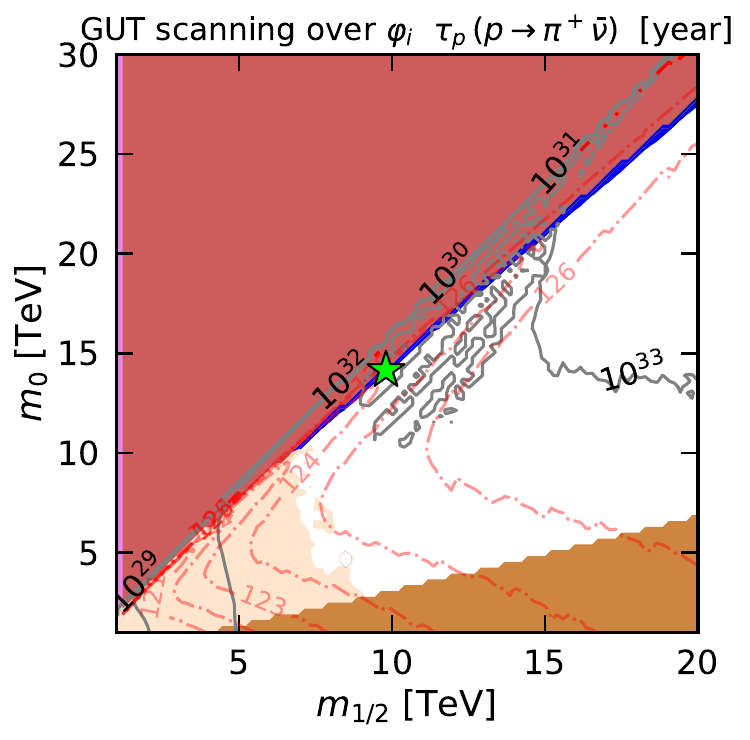}
    \vspace{-2mm}
    \includegraphics[width=0.35\linewidth]{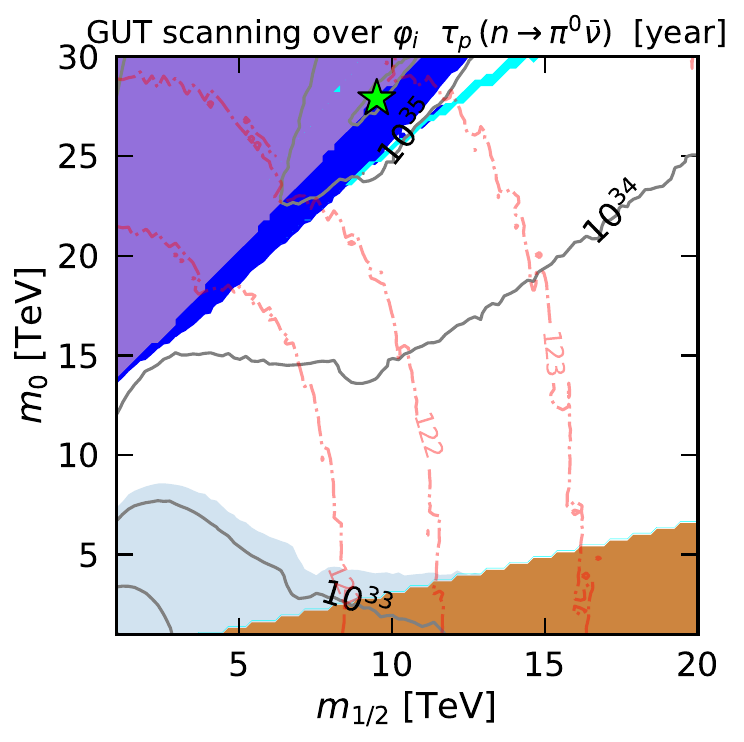}
    \vspace{-4mm}
    \includegraphics[width=0.35\linewidth]{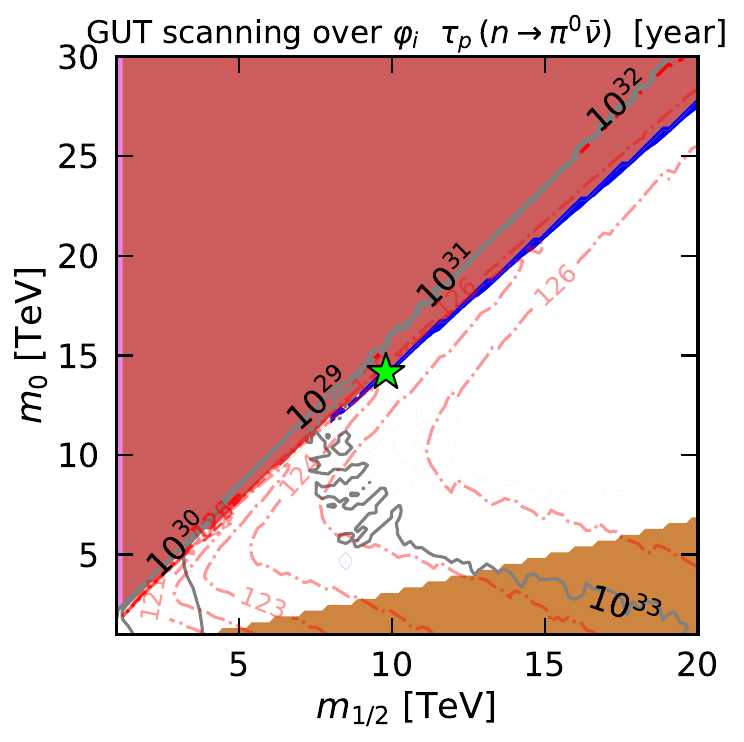}
    \vspace{-4mm}
    \includegraphics[width=0.6\linewidth]{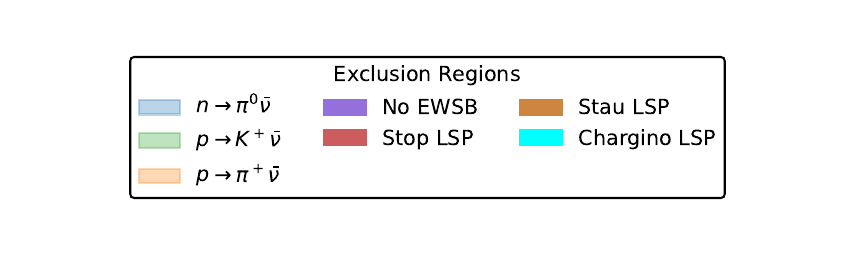}
    \hspace{-5cm}
    \vspace{-4mm}
    \includegraphics[width=0.6\linewidth]{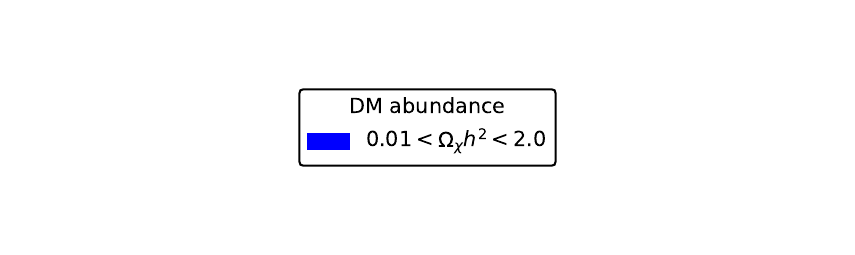}
    \hspace{-3cm}
    \caption{Proton lifetimes in ($m_{1/2}, m_0$) planes. Left panel:  $\lambda=\lambda'=1$, $\tan{\beta}=4$, $A_0/m_0=0$, $(c_{\Delta h,2})_i=(c_{\Delta h,3})_i=(c_{\Delta h,4})_i=0$. The green star is Benchmark point 1, and is used for determining the values of the $(c_{\Delta h,1})_i$, which are then also used across the plane. Right panel: $\lambda=\lambda'=1$, $\tan{\beta}=5$, $A_0/m_0=3$,  $(c_{\Delta h,2})_i=(c_{\Delta h,3})_i=(c_{\Delta h,4})_i=0$. The values of $(c_{\Delta h,1})_i$ are determined using the parameters of Benchmark point 2 shown by the green star. 
    The red dot-dashed lines are contours of the Higgs mass calculated using {\tt FeynHiggs~2.18.1}.
    The labelled dark solid lines show the contours of the nucleon lifetime. 
    }
    \label{fig:m0m12}
\end{figure}

In the right panels of Fig.~\ref{fig:m0m12} we use the spectrum of Benchmark point 2 to determine the coefficients of the Planck-suppressed operators. In this case, the LSP is a Bino and its relic density is in the preferred range (in the blue shaded strip) when the Bino mass is nearly degenerate with the lighter stop mass. As for the left panels, we set $(c_{\Delta h,2})_i=(c_{\Delta h,3})_i=(c_{\Delta h,4})_i=0$ across the plane. 
As mentioned previously, the Higgsino mass of Benchmark point 2 is relatively large.  This means that, if we suppress only the Wino-mediated process, the proton lifetime will never be long enough to exceed the current experimental bounds without taking $\lambda$ very large or $\lambda^\prime$ very small. However, by suppressing the first- and second-generation contributions to the Wino-mediated proton decay channels using $(c_{\Delta h, 1})_{1,2}$, the GUT phases $\varphi_i$ can then be used to enforce a cancellation between the Higgsino- and Wino-exchange processes for the third-generation contribution. As can be seen in the right panels of Fig.~\ref{fig:m0m12}, after scanning over the phases, this cancellation between the Wino- and Higgsino-mediated processes ensures viability for Benchmark point 2, even when $\lambda=\lambda^\prime=1$.

As in the left panels, the proton decay channel $p \to K^+ {\bar \nu}$ provides the strongest constraint on the Benchmark point 2-related planes seen in the right panels of Fig.~\ref{fig:m0m12}. Other than a nearly vertical region in the upper right panel that includes Benchmark point 2 shown by the green star, all of the $(m_{1/2}, m_0)$ plane is excluded over the range shown. Since the proton lifetime is suppressed by the soft supersymmetry-breaking masses, for large enough $m_{1/2}$ and $m_0$ the proton lifetime will again exceed the current experimental limits.  Note that some of the vertical strip allowed by $p \to K^+ {\bar \nu}$ is excluded by $n \to \pi^0 {\bar \nu}$, though this does not affect the blue strip that includes the green star. We also note that in the area below the blue strip the relic density is too large if the LSP is stable.

\subsection{Yukawa sector in the NUHM}
\label{sec:resultsNUHM}

Eq.~\eqref{eq:CHCW} shows that a small Higgsino mass suppresses the Higgsino-mediated proton decay contribution. To explore this possibility in more detail, in this subsection we consider the non-universal Higgs mass (NUHM) supersymmetry-breaking scenario~\cite{Ellis:2002iu,Ellis:2002wv}. We recall that the NUHM is similar to the CMSSM but with a less constrained Higgs sector. In the CMSSM, the soft masses of the two Higgs doublets satisfy $m_{H_1}=m_{H_2}=m_0$ at the GUT scale, and the Higgs mixing parameter $\mu$ and the pseudoscalar Higgs mass $m_A$ are determined by the electroweak symmetry-breaking conditions. In contrast, in the NUHM $\mu$ and $m_A$ are considered free parameters.

We consider two benchmark points that are variants of point 1 to illustrate how proton decay is affected by the supersymmetry-breaking parameters in conjunction with the GUT parameters. In both cases, we fix $m_A = m_0$ (which is {\em not} equivalent to fixing either $m_{H_1}$ or $m_{H_2}$) and allow $\mu$ to be free. We take:
\begin{itemize}
    \item Point 1a: $m_0=m_A=27.9$~TeV, $m_{1/2}=9.5$~TeV, $A_0=0$, $\tan \beta=6$, $\mu>0$. $\lambda=\lambda'=1$, $\varphi_i=0$.
    \item Point 1b: $m_0=m_A=27.9$~TeV, $m_{1/2}=9.5$~TeV, $A_0=0$, $\tan \beta=6$, $\mu>0$. $\lambda=\lambda'=3$, $\varphi_i=0$.
\end{itemize}
The left panels of Fig.~\ref{fig:Higgsinomass_fql} show results based on Benchmark point 1a. We use a flavor-universal suppression $(r_{ql})_i=r_{ql}$ to suppress Wino-mediated proton decays and explore the role of the Higgsino mass parameter $\mu$ in suppressing the Higgsino-mediated contributions. The GUT mediator masses are $M_X\approx 0.90\times 10^{16}~{\rm GeV}$ and $M_{H_C}\approx 1.3\times 10^{16}~{\rm GeV}$ and are largely independent of $\mu$. The inputs, mass spectra and observables for these two points are shown in Tables~\ref{tab:point1a} and~\ref{tab:point1b} with $\mu = 1$~TeV.

\begin{table}[ht]
  \centering
  \begin{tabular}{cccc}
  \hline \hline
  \multicolumn{4}{c}{Inputs}\\
  \hline
  $m_{1/2}=9.5$~TeV & $m_0=27.9$~TeV & $m_A=27.9$~TeV & $A_0/m_0=0$ \\
  $\mu=1~$TeV & $\tan{\beta}=6$ & $\lambda=1$ & $\lambda'=1$  \\
  \hline
  \multicolumn{4}{c}{GUT-scale parameters (masses in units of $10^{16}~$GeV)}\\
  \hline
  $M_{\rm GUT}=0.810$& $M_{H_C}=1.324$ & $M_\Sigma=0.662$ & $M_X=0.907$ \\
  $V=0.265$ & $g_5=0.685$ & $c=-1.23\times 10^{-6}$ \\
  \hline
  \multicolumn{4}{c}{MSSM parameters (masses in units of TeV)}\\
  \hline
  $m_\chi=1.060$ & $m_{\tilde{t}_1}=22.1$ & $m_{\tilde{g}}=19.40$ & $m_{\chi_2}=1.061$ \\
  $m_{A}=27.90$ 
  & $\mu=1.0$ & $m_{\tilde{\ell}_L}=28.4$& $m_{\tilde{\ell}_R}=28.1$ \\
  $m_{\tilde{\tau}_1}=28.0$ & $m_{\tilde{q}_L}=31.2$ & $m_{\tilde{d}_R}=30.9$ & $m_{\tilde{t}_2}=27.3$ \\
  $A_t=18.75$ & $A_b=27.2$ & $B=-131$ & \\
  \hline
  \multicolumn{4}{c}{Observables}\\
  \hline
  $\Omega_\chi h^2=0.110$ & $m_h=125.8$~GeV & & \\
  \hline \hline
  \end{tabular}
  \caption{Benchmark point 1a parameters with $\mu=1~{\rm TeV}$. The observables in the last line are the dark matter relic density and Higgs mass calculated with {\tt FeynHiggs 2.18.1}. 
  }
  \label{tab:point1a}
\end{table}

\begin{table}[ht]
  \centering
  \begin{tabular}{cccc}
  \hline \hline
  \multicolumn{4}{c}{Inputs}\\
  \hline
  $m_{1/2}=9.5$~TeV & $m_0=27.9$~TeV & $m_A=27.9$~TeV & $A_0/m_0=0$ \\
  $\mu=1~$TeV & $\tan{\beta}=6$ & $\lambda=3$ & $\lambda'=3$  \\
  \hline
  \multicolumn{4}{c}{GUT-scale parameters (masses in units of $10^{16}~$GeV)}\\
  \hline
  $M_{\rm GUT}=0.810$& $M_{H_C}=2.776$ & $M_\Sigma=1.388$ & $M_X=0.626$ \\
  $V=0.185$ & $g_5=0.677$ & $c=-1.76\times 10^{-6}$\\   
  \hline \hline
  \end{tabular}
  \caption{Inputs and GUT-scale parameters for Benchmark point 1b.
  The MSSM parameters and the observables have the same values as those for Benchmark point 1a shown in Table~\ref{tab:point1a}.}
  \label{tab:point1b}
\end{table}

\begin{figure}[ht]
    \centering
    \includegraphics[width=0.45\linewidth]{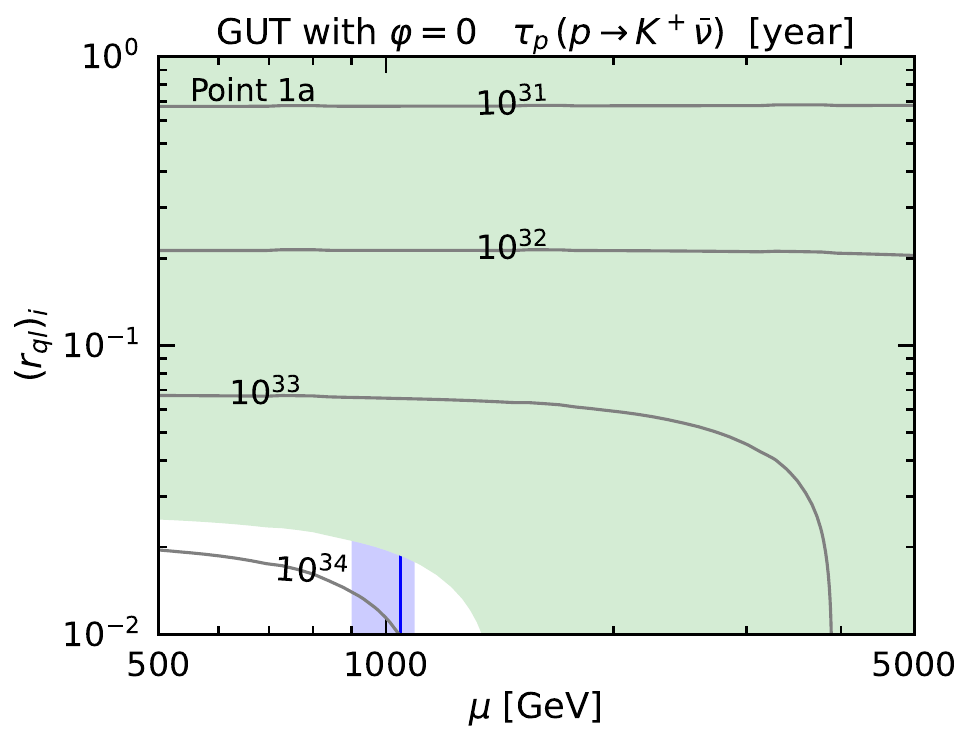}
    \includegraphics[width=0.45\linewidth]{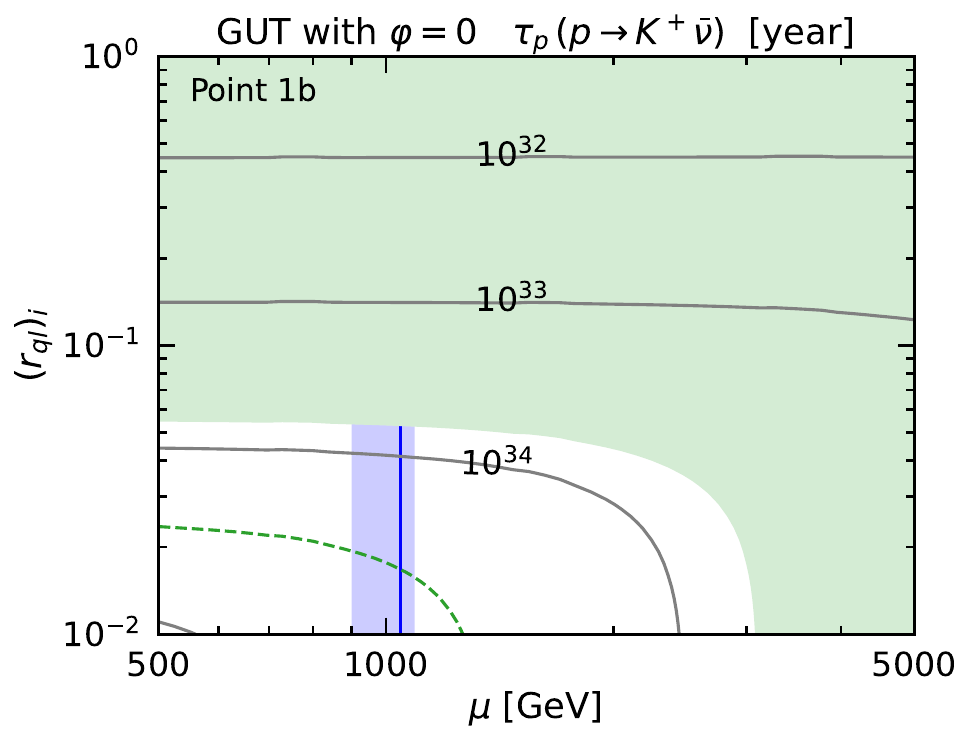}
    \includegraphics[width=0.45\linewidth]{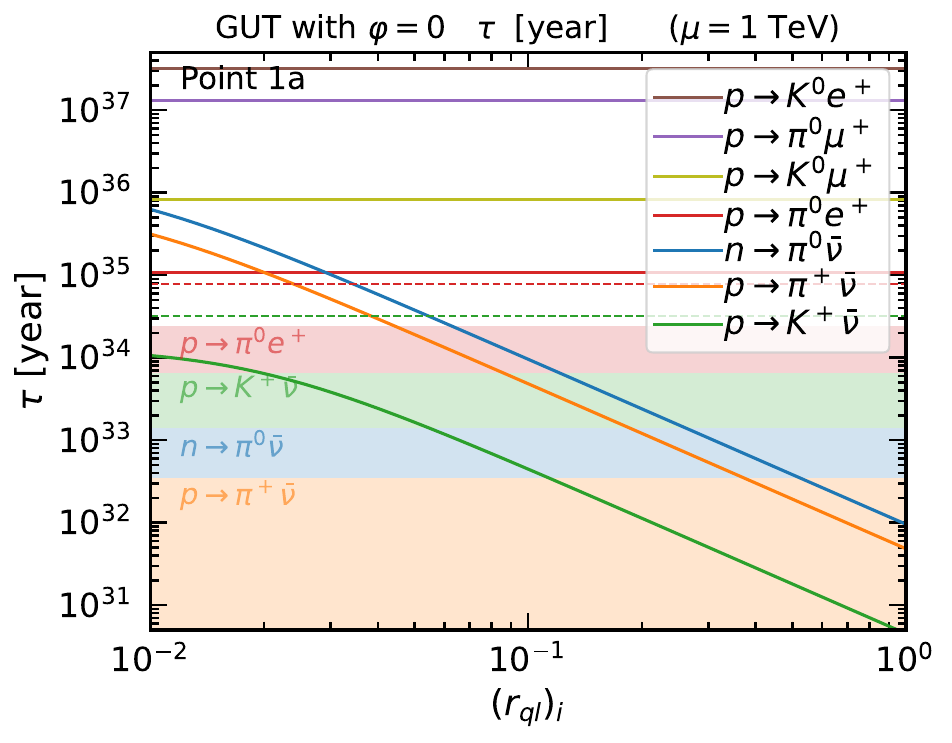}
    \includegraphics[width=0.45\linewidth]{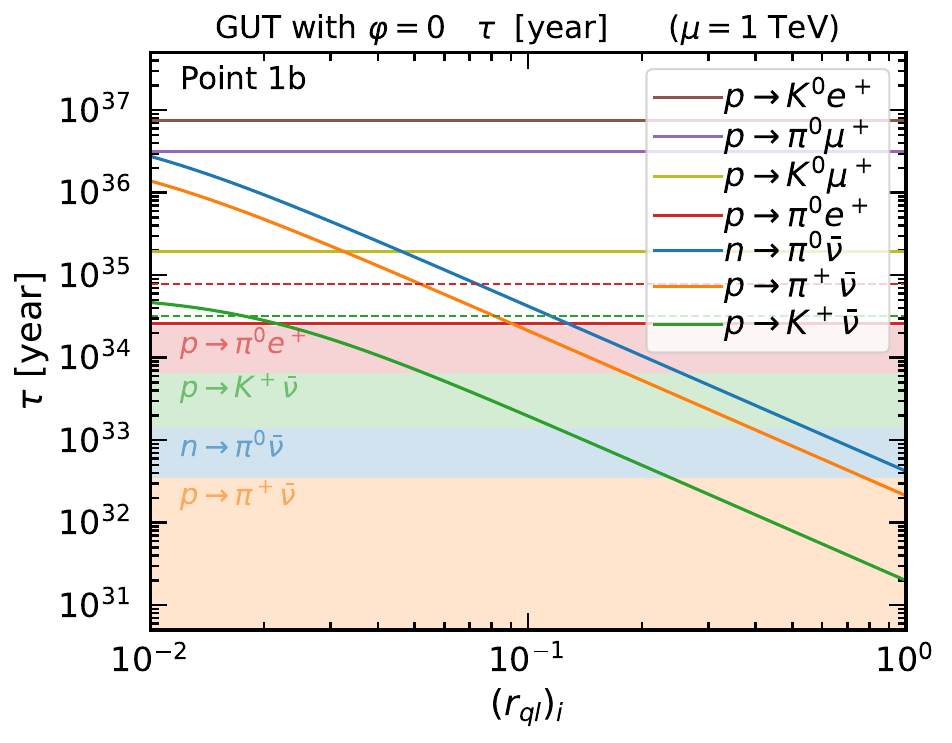}
    \caption{Proton lifetimes in NUHM models, assuming vanishing GUT phases, $\varphi_i=0$. The left panels are for Benchmark point 1a defined in the text, and the right panels are for Benchmark point 1b. The shaded areas are the current proton decay limits, and the dashed lines are the Hyper-Kamiokande sensitivities.}
    \label{fig:Higgsinomass_fql}
\end{figure}

In the upper left panel of Fig.~\ref{fig:Higgsinomass_fql},  we show contours of the proton lifetime in the $(\mu, (r_{ql})_i)$ plane.
The vertical blue line shows where $\Omega_\chi h^2 =0.12$, and $0.09<\Omega_\chi h^2 <0.13$ in the blue shaded region. As one can see, for this choice $\lambda, \lambda'$ the $p \to K^+ {\bar \nu}$ channel place a limit $\mu \lesssim 1.3$~TeV and $r_{ql} \lesssim 0.025$ for low $\mu$.
The Higgs mass for Benchmark points 1a and 1b is $m_h = 125.8$~GeV. 
In the  lower left panel we fix $\mu = 1$~TeV and show the calculated proton lifetime in several channels compared with the existing experimental constraints shown by the shaded regions, which are as labeled. The future Hyper-Kamiokande sensitivities are shown by the dotted line. Once again, we see that the  $p \to K^+ {\bar \nu}$ channel provides the strongest constraint $r_{ql} \lesssim 0.02$ (for this choice of $\mu$).

Benchmark point 1b differs from Benchmark point 1a only in the GUT parameters, $\lambda$ and $\lambda'$ that affect $M_X$ and $M_{H_C}$, as seen in Table~\ref{tab:point1b}.  By taking larger values of $\lambda$ and $\lambda'$, we simultaneously enhance dimension-6 proton decay and suppress dimension-5 proton decay due to how $M_X$ and $M_{H_C}$ scale with these couplings, see Eq.~(\ref{eq:MG2}). 
The right panels of Fig.~\ref{fig:Higgsinomass_fql} illustrate results for the Benchmark point 1b for $\lambda=\lambda'=3$, which has a GUT gauge boson mass $M_X\approx 0.6\times 10^{16}~{\rm GeV}$ and color-triplet mass $M_{H_C}\approx 2.8\times 10^{16}~{\rm GeV}$ with only a small dependence on $\mu$. As in the corresponding plot for Benchmark point 1a, the vertical blue line shows where $\Omega_\chi h^2 =0.12$, and $0.09<\Omega_\chi h^2 <0.13$ in the blue shaded region.
In the upper right panel, we see that the limits on both $\mu$ and $r_{ql}$ are somewhat relaxed with the larger values of $\lambda$ and $\lambda'$.  In this case, for low $r_{ql}$, we find $\mu < 3$~TeV, though if the LSP is stable, we must be on or to the left of the vertical blue line, which again shows the value of $\mu$ for a relic density of $\Omega_\chi h^2 \simeq 0.1$. The limit on $r_{ql}$ for low $\mu$ is now $r_{ql} \lesssim 0.058$.

The proton lifetime for several decay channels for Benchmark point 1b is shown in the lower right panel of Fig.~\ref{fig:Higgsinomass_fql}.
We note that in this case the predicted lifetime of $\tau(p\to \pi^0e^+)=2.6\times 10^{34}~{\rm years}$ for the dimension-6 $p \to \pi^0 e^+$ mode is very close to the present limit, and within the projected Hyper-Kamiokande sensitivity. Benchmark point 1b is therefore an interesting case where both the dimension-5 decay mode $p \to K^+ \bar{\nu}$ and the dimension-6 decay mode $p \to \pi^0 e^+$ may be detectable in the near future.

\section{Conclusions and Discussion}
\label{sec:conclusion}

Improved sensitivities to proton decay modes are projected for the new generation of neutrino and proton decay detectors such as Hyper-Kamiokande and JUNO, which will enable the exploration of extended regions of the parameter spaces of GUT models.

We have considered in this paper supersymmetric SU(5) GUT models with the inclusion of dimension-5 Planck-suppressed operators at the GUT scale, and investigated their possible effects on the mass spectrum and proton decays. These effects can be classified into two types. The first originates from the Higgs sector, where Planck-suppressed operators induce different masses for the GUT bosons $M_{\Sigma_3}$ and $M_{\Sigma_8}$.  This modification alters the gauge coupling matching conditions and GUT mass spectrum, thereby affecting proton decays. The second effect comes from the Yukawa sector. Once dimension-5 Planck-suppressed operators are included, the colored Higgs Yukawa couplings deviate from the MSSM Yukawa couplings. These effects are parameterized by $(c_{\Delta h, 1})_i$, $(c_{\Delta h, 2})_i$, $(c_{\Delta h, 3})_i$ and $(c_{\Delta h, 4})_i$, among which $(c_{\Delta h, 1})_i$ has the most significant impact on proton decay modes.

We have performed a numerical computation of dimension-5 and 6 proton decays in illustrative CMSSM and NUHM scenarios. Planck-suppressed operators in the Higgs sector can modify the GUT mass spectra and suppress proton decays by an overall factor of $ {\cal O}(10^{-1}-10^{-3})$. For the GUT couplings $\lambda^\prime \ll \kappa_1 \lesssim 10^{-3}$ it is possible to evade current experimental bounds.

The Yukawa-sector coefficient  $(c_{\Delta h, 1})_i$ enters directly into the Wilson coefficients of proton decays at the GUT scale and can compete with the MSSM Yukawa couplings. 
In previous work~\cite{Ellis:2015rya,Ellis:2016tjc,Ellis:2017djk,Ellis:2019fwf,Ellis:2020mno,Evans:2021hyx}, not all possibilities for the Planck-suppressed operators were considered, and $f_{QL}=f_d$ was assumed at the GUT scale, leading to an apparent suppression of the proton lifetime.
In this paper, splitting in the lepton and down-type quark Yukawa couplings is introduced via $c_{\Delta h, 2}$, and the colored-Higgs Yukawa $f_{QL}$ is the combination of lepton Yukawa and Planck-suppressed coupling, with $f_{QL}=f_e+\frac{5}{\sqrt{2}}(c_{\Delta h, 1})R$.
We find that the value of $(c_{\Delta h, 1})_i$ can change drastically dimension-5 proton decays, suppressing or even cancelling the Wino-exchange contribution, which is typically dominant in conventional supersymmetric SU(5) models.  
Furthermore, after including the unknown GUT phases $\varphi_i$, interference between Higgsino- and Wino-mediated processes can occur, leading to strong suppressions of nucleon decays. 
We have identified the viable parameter space in the CMSSM framework and computed relevant observables including the observed Higgs mass and the dark matter relic density~\footnote{\SH{In the CMSSM, the requirement of reproducing the observed Higgs mass strongly constrains the supersymmetry breaking scale. As a result, we cannot obtain a viable parameter space merely by raising the supersymmetry breaking scale.}}.
Even for supersymmetry breaking at the ${\cal O}(10~{\rm TeV})$ scale, there is still viable CMSSM parameter space with GUT couplings $\lambda=\lambda'=1$.
Alternatively, a small Higgsino mass can suppress the Higgsino-exchange contributions to nucleon decays. We have illustrated this point by presenting the suppression required in the NUHM framework to obtain a viable region of parameter space. 
As we have emphasized, in NUHM there is the possibility that both the dimension-5 decay mode $p \to K^+ \bar{\nu}$ (at Hyper-Kamiokande, JUNO, and DUNE) and the dimension-6 decay mode $p \to \pi^0 e^+$ (at Hyper-Kamiokande) may be detectable.

In summary, we have shown that the inclusion of dimension-5 Planck-suppressed operators can significantly modify proton decay modes compared to those in conventional supersymmetric SU(5) GUTs. For certain regions of parameter space, the decay mode $p \to K^+\bar{\nu}$ is no longer dominant. 
Proton decays can be suppressed through several independent mechanisms once Planck-suppressed operators are included, notably the effects from the Higgs sector, the Yukawa sector, unknown phases $\varphi_i$, and the Higgsino mass. Therefore, the absence of proton decay signals in current experiments does not disfavor supersymmetric SU(5) GUTs.
We compared our results with the current Super-Kamiokande proton limits and the prospective sensitivities of Hyper-Kamiokande. Our analysis shows that regions of the parameter space of supersymmetric SU(5) GUTs previously disfavored by proton decay constraints can be reopened once Planck-suppressed effects are taken into account, and we have shown how future proton decay experiments can  explore further the allowed parameter space, and possibly discover more than one proton decay mode.

\section*{Acknowledgements}

The work of J.E. was supported partly by the United Kingdom STFC Grant ST/T000759/1. 
The work of J.L.E. was supported partly by the National Natural
Science Foundation of China under grant No. W2432002.
The work of S.H. was supported in part by J.L.E. and Prof. Yuichiro Nakai, funded by the National Natural Science Foundation of China.
The work of N.N. was supported in part by the Grant-in-Aid for Scientific Research C (No.~25K07314).  
The work of K.A.O. was supported partly
by the DOE grant DE-SC0011842 at the University of Minnesota.


\bibliographystyle{utphysmod}
\bibliography{ref}


\end{document}